\documentclass[final,1p]{elsarticle}

\usepackage{framed,multirow}
\usepackage{url}
\usepackage{pythonhighlight} 

\usepackage{amssymb}
\usepackage{amsmath}

\journal{Journal of Computational Physics}

\renewcommand{\vec}[1]{\mathbf{#1}}
\newcommand{\projectednormal}{\vec{n}_\text{p}}
\newcommand{\weak}[2]{\left(#1,#2\right)}
\newcommand{\iweak}[2]{\left\langle#1,#2\right\rangle}
\newcommand{\iiweak}[2]{\left[#1,#2\right]}
\newcommand{\software}[1]{\textsc{#1}}

\newcommand{\augresvec}{\vec{R}^\text{aug}}
\newcommand{\resvec}{\vec{R}^{*}}
\newcommand{\rescompo}{R}
\newcommand{\resfunc}{\mathcal{R}^{*}}
\newcommand{\resfuncazi}{\mathcal{R}^{*,m}}
\newcommand{\doffunc}{\mathcal{U}}
\newcommand{\doffuncaxi}{\mathcal{U}^0}
\newcommand{\doffuncazi}{\mathcal{V}^m}

\newcommand{\aziangle}{\varphi}
\newcommand{\dofcompo}{U}
\newcommand{\dofvec}{\vec{U}}
\newcommand{\augdofvec}{\vec{U}^\text{aug}}
\newcommand{\guessvec}{\vec{U}^\text{guess}}
\newcommand{\dotdofcompo}{\dot{U}}
\newcommand{\dotdofvec}{\dot{\vec{U}}}
\newcommand{\statresvec}{\vec{R}}
\newcommand{\statdofvec}{\vec{U}^0}
\newcommand{\eigenvec}{\vec{V}}
\newcommand{\eigenval}{\lambda}

\newcommand{\jacmat}{\mathbf{J}}
\newcommand{\augjacmat}{\mathbf{J}^\text{aug}}
\newcommand{\massmat}{\mathbf{M}}
\newcommand{\massop}{\mathcal{M}}
\newcommand{\jacop}{\mathcal{J}}
\newcommand{\jaccompo}{J}
\newcommand{\masscompo}{M}
\newcommand{\hesstens}{\mathbf{H}}
\newcommand{\hesscompo}{H}
\newcommand{\domeigen}{\lambda_\text{c}}
\newcommand{\bifparam}{p}
\newcommand{\critparam}{p_\text{c}}
\newcommand{\covarcoorddiff}{G} 
\newcommand{\contracoorddiff}{G}
\newcommand{\Bo}{\operatorname{Bo}}
\newcommand{\rcontactline}{r_\text{cl}}
\newcommand{\posnode}[1]{{\bar{#1}}}
\newcommand{\othernode}[1]{{#1}}
\newcommand{\shapefuncother}{\psi}
\newcommand{\shapefuncpos}{\bar\psi}
\newcommand{\testfunc}{\chi}
\newcommand{\bulkdomain}{\Omega}
\newcommand{\interfdomain}{\Gamma}
\newcommand{\bulkdomainh}{\Omega^h}
\newcommand{\interfdomainh}{\Gamma^h}
\newcommand{\bulkelem}{\mathcal{T}}
\newcommand{\interfelem}{\mathcal{I}}
\newcommand{\unitnormal}{\mathbf{\hat{n}}}
\newcommand{\unitnormalcompo}{\hat{n}}
\newcommand{\unitvec}{\hat{\vec{e}}}
\newcommand{\posvec}{\vec{x}}
\newcommand{\posvecaxi}{\posvec^0}
\newcommand{\posvecazi}{\posvec^m}

\newcommand{\normalvecaxi}{\unitnormal^0}
\newcommand{\normalvecazi}{\vec{n}^m}
\newcommand{\normalcompoaxi}{\hat{n}^0}

\newcommand{\mycomment}[1]{}

\newcommand*{\myinputpython}[3]{\lstinputlisting[firstline=#2,lastline=#3,firstnumber=#2,frame=single,breakindent=.5\textwidth,frame=single,breaklines=true,style=mypython]{#1}}

\begin{document}

\begin{frontmatter}

\title{Bifurcation tracking on moving meshes and with consideration of azimuthal symmetry breaking instabilities}


\author[1]{Christian Diddens\corref{cor1}}
\cortext[cor1]{Corresponding author: 
  c.diddens@utwente.nl}
  
\author[1]{Duarte Rocha\fnref{fn1}}
\fntext[fn1]{d.rocha@utwente.nl}

\affiliation[1]{organization={Physics of Fluids group, Department of Science and Technology, Mesa+ Institute, Max Planck Center for Complex Fluid Dynamics and 
J. M. Burgers Centre for Fluid Dynamics, University of Twente},
            addressline={P.O. Box 217}, 
            city={Enschede},
            postcode={7500 AE}, 
            country={The Netherlands}}

\begin{abstract}
We present a black-box method to numerically investigate the linear stability of arbitrary multi-physics problems. While the user just has to enter the system's residual in weak formulation, i.e. by a finite element method, all required discretized matrices are automatically assembled based on just-in-time generated and compiled highly performant C code. Based on this method, entire phase diagrams in the parameter space can be obtained by bifurcation tracking and continuation within minutes. Particular focus is put on problems with moving domains, e.g. free surface problems in fluid dynamics, since a moving mesh introduces a plethora of complicated nonlinearities to the system. By symbolic differentiation before the code generation, however, these moving mesh problems are made accessible to bifurcation tracking methods. In a second step, our method is generalized to investigate symmetry-breaking instabilities of axisymmetric stationary solutions by effectively utilizing the symmetry of the base state. Each bifurcation type is validated on the basis of results reported in the literature on versatile fluid dynamics problems.
\end{abstract}




\end{frontmatter}


\section{Introduction}
\label{sec:introduction}
Many physical systems exhibit spontaneous symmetry breaking or fundamental changes in behavior upon subtle changes in a control parameter. Classical examples are e.g. the onset of convection in a Rayleigh-Bénard system, any self-organized pattern formation \cite{Cross1993} as e.g. the Turing instability \cite{Turing1952}, neuronal dynamics as e.g. the FitzHugh-Nagumo model \cite{FitzHugh1961} and more.

Despite of the complexity and the inherent nonlinearities in these systems, deep insight into the dynamics can be obtained by investigating the linear dynamics around the stationary states close to a critical parameter threshold, i.e. by a linear stability analysis. Thereby, one can calculate at which critical parameter value the system undergoes a transition and additionally obtain the corresponding eigenfunction, from what further understanding of the intrinsic dynamics close to the instability can be extracted. This approach can be applied on systems that can be modeled by a system of ordinary differential equations as well as system continuous in space, i.e. described by partial differential equations. In particular in the latter case, however, an analytical treatment of the linear stability analysis is often hampered by the nonlinearities, which generically prevent exact analytical solutions for nontrivial stationary solutions. Furthermore, a nontrivial geometry of the system also hampers the applicability of analytical methods.

For these cases, one has to fall back to the numerical solution of the stationary solution and the corresponding eigenvalue problem to find the dominant eigenvalue $\domeigen$, i.e. the eigenvalue with the highest real part, and subsequently determine the corresponding critical parameter value when its real part becomes zero, i.e. the bifurcation point at which $\operatorname{Re}\domeigen=0$ holds. This can be achieved rather quickly by the bisection method, however, this method fails e.g. for fold bifurcations, since the stationary solution branch ceases to exists beyond the critical parameter. Instead, one can virtually jump directly on the bifurcation point by solving an augmented system of equations to simultaneously solve for the stationary solution, the critical parameter value and the corresponding critical eigenfunction. Once achieved, the bifurcation branch can be tracked along another parameter, e.g. by (pseudo-)arclength continuation. This approach is known as bifurcation tracking and it or related methods have been implemented in a variety of software packages, e.g. \software{pde2path} \cite{Uecker2014}, \software{AUTO-07p} \cite{Doedel1998}, \software{oomph-lib} \cite{Heil2006} or \software{LOCA} \cite{Salinger2005}. A detailed overview on numerical stability and bifurcation analysis, with particular focus on fluid dynamics, can be found in Ref. \cite{Salinger2014}.

From the user perspective, it is desired to provide a simple interface to enter the system of equations to be investigated and subsequently automatize the bifurcation analysis in a black-box manner. However, the assembly of the required Jacobian and mass matrices for the eigenproblem and the bifurcation tracking can be challenging, since these matrices must be treated in a monolithic manner to consider all intrinsic couplings of the problem to obtain the exact eigenproblem and its solution. In particular, if moving domains are considered as e.g. in fluid-structure interaction or free surface flows implemented in an sharp-interface moving mesh approach, the influence on the moving geometry on the discretized equations has to be considered as well. Calculating the required first and second order derivatives of the discretized equations with respect to the mesh coordinates by finite difference is in principle possible, but computationally expensive and usually not sufficiently accurate to ensure convergence of the solution. Similar aspects are relevant in shape optimization problems, for which recently automatic differentiation approaches have been successfully implemented \cite{Ham2019,Gangl2021}.

In this article, we present a black-box bifurcation tracking method for arbitrary multi-physics problems on moving meshes. By using symbolical differentiation, the monolithic Jacobian, mass matrix, parameter derivatives thereof, and the Hessian are assembled by highly performant C codes, which are generated in a just-in-time manner from the entered system of equations. The system of equations is entered by the user in  weak formulation, which is subsequently discretized by the finite element method. For all considered bifurcation types, we present a representative example and validate our implementation on the basis of literature results. 

Since the method can still be too expensive for full three-dimensional settings, we also present a reduction of the dimensions for stationary solutions that possess a symmetry. In particular, we show how the method can be applied to assess azimuthal symmetry breaking bifurcations of axisymmetric base states at very low computational costs. 

The article is structured as follows: In section~\ref{sec:matdesc}, the general mathematical notation as well as the used bifurcation tracking systems are introduced. In section~\ref{sec:spatdisc}, the complications for bifurcation tracking on moving meshes are discussed, followed by details on our particular implementation in section~\ref{sec:implem}. After validation of our approach in section~\ref{sec:validation}, the generalization to the azimuthal stability analysis of axisymmetric base states is discussed and validated in section~\ref{sec:azibreak}. The article ends with a conclusion, including potential applications of this method.

\section{Mathematical description}
\label{sec:matdesc}
\subsection{Notation}
Once any system of equations is discretized in space, a system of ordinary differential equations as function of time is obtained, which can be most generally written in the implicit form
\begin{align}
\resvec(\dofvec,\dotdofvec,\bifparam)=\vec{0}\,.
\end{align}
Here, $\dofvec$ comprises the $N$ spatially discretized real-valued unknowns of the system, $\dotdofvec$ is the first order time derivative and the residual vector $\resvec$ is a nonlinear real-valued function with $N$ components, describing the coupled temporal dynamics. $\resvec$ might depend on further parameters, but it is sufficient to focus on a single parameter $\bifparam$ here and keep potential further parameters fixed. Higher order time derivatives can be considered by introducing auxiliary components in $\dofvec$ and modifying $\resvec$ accordingly, i.e. the typical conversion of higher order ordinary differential equations to a first order system.

If it exists at the given parameter $\bifparam$, a stationary solution $\statdofvec$ is obtained by solving the stationary residual
\begin{align}
\statresvec(\dofvec,\bifparam)=\resvec(\dofvec,\vec{0},\bifparam)=\vec{0}\,.
\end{align}
for $\dofvec$. Since $\statresvec$ can be nonlinear, Newton's method provides a good approach, where one iteratively updates a reasonable guess $\guessvec$ for $\dofvec$ by
\begin{align}
\guessvec\hookleftarrow \guessvec-\jacmat^{-1}(\guessvec,p)\statresvec(\guessvec,p) \label{eq:newton}
\end{align}
until the maximum norm $\|\statresvec(\guessvec,p)\|_\infty$ falls below a given threshold. Here, $\jacmat$ is the Jacobian, i.e. the derivatives of the stationary residual $\statresvec$ with respect to $\dofvec$.

Once the stationary solution $\statdofvec$ has been found, the linear stability can be investigated by the corresponding generalized eigenvalue problem
\begin{align}
\lambda \left.\massmat\right|_{(\statdofvec,p)}\eigenvec=-\left.\jacmat\right|_{(\statdofvec,p)}\eigenvec\,. \label{eq:geneigen}
\end{align}
Here, the mass matrix $\massmat$ and the $\jacmat$ are the $N\times N$ matrices, i.e. 
\begin{align}
\masscompo_{ij}=\left.\frac{\partial \rescompo^*_i}{\partial\dotdofcompo_j}\right|_{\dotdofvec=\vec{0}}\qquad\text{and}\qquad \jaccompo_{ij}=\frac{\partial \rescompo_i}{\partial\dofcompo_j}\,.
\end{align}
For completeness, we also introduce the Hessian $\hesstens$, which is required in the bifurcation tracking later on:
\begin{align}
\hesscompo_{ijk}=\frac{\partial \jaccompo_{ij}}{\partial\dofcompo_k}=\frac{\partial \rescompo_i}{\partial\dofcompo_j \partial\dofcompo_k}\,.
\end{align}
A bifurcation happens at a critical parameter $\bifparam=\critparam$, when the generalized eigenvalue problem \eqref{eq:geneigen} has a solution for an eigenvalue $\eigenval$ with corresponding eigenvector $\eigenvec$ for which $\operatorname{Re}\eigenval$ crosses zero.

\subsection{Bifurcation tracking}
\label{sec:biftrack}
Instead of iteratively solving the generalized eigenvalue problem \eqref{eq:geneigen} and adjusting the parameter $\bifparam$ until $\operatorname{Re}\eigenval=0$ holds, e.g. by bisection, bifurcation tracking solves for the critical parameter $\critparam$, the corresponding eigenvector $\eigenvec$ and the potentially parameter-dependent stationary solution $\statdofvec$ simultaneously. This is achieved by the same steps as above, i.e. by solving Newton's method \eqref{eq:newton}, but the unknowns and the residual vector have to be augmented. Depending on the bifurcation's type, the particular augmentation must be chosen differently, which is elaborated in the following. The reader is also referred to Refs. \cite{Kuznetsov1998,Cliffe2000,Bindel2014,Umbria2016,Hazel2019} within this context.

\subsubsection{Fold bifurcation}
In a fold bifurcation (or saddle-node bifurcation), the stationary solution $\statdofvec$ ceases to exist at the critical parameter $\critparam$ and an unstable and a stable branch of $\statdofvec$ meet in the bifurcation point. The corresponding eigenvalue has no imaginary part and hence the eigenvalue problem \eqref{eq:geneigen} reduces to $\jacmat\eigenvec=0$. This equation is solved together with the original residual $\statresvec$. As a further constraint, the trivial solution $\eigenvec=0$ must be prevented, which can be enforced by solving additionally $\eigenvec\cdot\vec{C}=1$ for some nontrivial and non-orthogonal vector $\vec{C}$, e.g. the initial guess of $\eigenvec$. As additional unknowns for this $N+1$ additional equations, the eigenvector $\eigenvec$ and the bifurcation parameter $\bifparam$ are used. Hence, the original system is augmented as follows \cite{Moore1980}:
\begin{align}
\augdofvec=\begin{pmatrix} \dofvec \\ \eigenvec \\ p \end{pmatrix}\qquad \augresvec=\begin{pmatrix} \statresvec \\ \jacmat\eigenvec \\ \eigenvec\cdot\vec{C}-1\end{pmatrix} \qquad \augjacmat=\begin{pmatrix} \jacmat & \mathbf{0} & \partial_p\statresvec \\ \hesstens\eigenvec & \jacmat & (\partial_p \jacmat)\eigenvec \\ \vec{0} & \vec{C} & 0  \end{pmatrix} \,.
\end{align}
After solving this system with a good starting guess, e.g. providing a normalized eigenvector solution near the bifurcation from the generalized eigenvalue problem \eqref{eq:geneigen}, with the Newton method \eqref{eq:newton}, one obtains the critical stationary solution $\statdofvec$, the corresponding eigenvector $\eigenvec$ and the critical parameter $\critparam$, i.e. the location of the fold bifurcation.

\subsubsection{Pitchfork bifurcation}
\label{sec:pitchtrack}
Pitchfork bifurcations generically appear in systems with assumed symmetry, where they are located at the intersection of a symmetry-preserving and a symmetry-breaking branch of solutions. At the bifurcation point, both of these branches exchange the stability. To track a pitchfork bifurcation, the following augmented system must be solved \cite{Salinger2005}:
\begin{align}
\augdofvec=\begin{pmatrix} \dofvec \\ \eigenvec \\ p \\ \epsilon   \end{pmatrix}\qquad \augresvec=\begin{pmatrix} \statresvec+\epsilon\vec{S} \\ \jacmat\eigenvec  \\ \eigenvec\cdot\vec{C}-1 \\ \dofvec\cdot\vec{S}\end{pmatrix} \qquad \augjacmat=\begin{pmatrix} \jacmat & \mathbf{0} & \partial_p\statresvec &\vec{S} \\ \hesstens\eigenvec & \jacmat & (\partial_p \jacmat)\eigenvec & \mathbf{0} \\ \vec{0} & \vec{C} & 0 & 0 \\ \vec{S} & \vec{0}  & 0 & 0 \end{pmatrix} \,. \label{eq:pitchtrack}
\end{align}
The unknown $\epsilon$ is a slack variable and $\vec{S}$ is an antisymmetric vector with respect to the assumed symmetry. $\vec{S}$ and $\vec{C}$ can be set to the initial guess for the eigenvector $\eigenvec$, which is obtained from an explicit solution of the eigenvalue problem \eqref{eq:geneigen} close to the bifurcation. Upon solution of this system, the eigenvector $\eigenvec$ is hence breaking the symmetry, i.e. is orthogonal to $\dofvec$ and hence the value of $\epsilon$ is small. If the mesh does not reflect the symmetry that is broken at the bifurcation, the discrete orthogonality relation $\dofvec\cdot\vec{S}=0$ hampers convergence. It can be replaced by a weak integral formulation of this product, i.e. by integrating over the product of finite element interpolations of $\dofvec$ and $\vec{S}$ in the continuous domain. Thereby, convergence on a mesh not complying with the symmetry can be achieved, even though it usually still converges poorly (cf. table~\ref{tab:convrates} later on). Improvements of this limitation by projection operators are discussed in Refs. \cite{Tavener1992,Cliffe2000}.

\subsubsection{Hopf bifurcation}
A system undergoes a Hopf bifurcation if, when a parameter crosses a critical threshold $\critparam$, the system becomes unstable in a self-excited oscillation. This bifurcation is associated with a pair of complex conjugated eigenvalues crossing $\operatorname{Re}\eigenval=0$ with a non-vanishing imaginary part. The eigenvalues and eigenvectors can therefore be written as $\eigenval=\eigenval_\mathrm{r}\pm i \eigenval_\mathrm{i}$ and $\eigenvec=\eigenvec_\mathrm{r}\pm i \eigenvec_\mathrm{i}$, respectively. A Hopf bifurcation hence generally occur when $\eigenval_\mathrm{r}=0$ and $\eigenval_\mathrm{i}\neq 0$ holds at the critical parameter $\critparam$. The augmented system will have $3N+2$ unknowns \cite{Griewank1983}, and it can be solved through:
\begin{gather}
\augdofvec=\begin{pmatrix} \dofvec \\ \eigenvec_\mathrm{r} \\ \eigenvec_\mathrm{i} \\ p \\ \eigenval_\mathrm{i} \end{pmatrix}\qquad 
\augresvec=\begin{pmatrix} \statresvec \\ -\jacmat\eigenvec_\mathrm{r} + \eigenval_\mathrm{i} \massmat \eigenvec_\mathrm{i} \\ -\jacmat\eigenvec_\mathrm{i} - \eigenval_\mathrm{i} \massmat \eigenvec_\mathrm{r} \\ \eigenvec_\mathrm{r}\cdot\vec{C}-1 \\ 
\eigenvec_\mathrm{i}\cdot\vec{C}\end{pmatrix}  \nonumber \\ 
\augjacmat=\begin{pmatrix} \jacmat & \mathbf{0} & \mathbf{0} & \partial_p\statresvec & \mathbf{0} \\ -\hesstens\eigenvec_\mathrm{r} + \eigenval_\mathrm{i} (\partial_\dofvec \massmat) \eigenvec_\mathrm{i} & \jacmat & \eigenval_\mathrm{i} \massmat & - (\partial_p \jacmat) \eigenvec_\mathrm{r} + \eigenval_\mathrm{i} (\partial_p \massmat) \eigenvec_\mathrm{i} & \massmat \eigenvec_\mathrm{i} \\ -\hesstens\eigenvec_\mathrm{i} - \eigenval_\mathrm{i} (\partial_\dofvec \massmat) \eigenvec_\mathrm{r} & \eigenval_\mathrm{i} \massmat & \jacmat & - (\partial_p \jacmat) \eigenvec_\mathrm{i} - \eigenval_\mathrm{i} (\partial_p \massmat) \eigenvec_\mathrm{r} & - \massmat \eigenvec_\mathrm{r} \\ 0 & \vec{C} & 0 & 0 & 0 \\ 0 & 0 & \vec{C} & 0 & 0 \end{pmatrix} \,.
\end{gather}

Once again, the constraint vector $\vec{C}$, preventing the trivial eigensolution $\eigenvec_\mathrm{r}=\eigenvec_\mathrm{i}=0$, can be the real part of the initial guess for the eigenvector $\eigenvec$. The second constraint, $\eigenvec_\mathrm{i}\cdot\vec{C}=0$, is required to select a unique phase of the complex-valued eigenvector. This equation is associated with the additional unknown Hopf frequency $\eigenval_\mathrm{i}$. Provided that the initial guess is near the bifurcation, the Newton method \eqref{eq:newton} is generally able to capture the Hopf bifurcation's critical stationary solution $\statdofvec$, its critical parameter $\critparam$, the corresponding real and imaginary parts of the eigenvector $\eigenvec$ and the corresponding imaginary eigenvalue $\eigenval_\mathrm{i}$.

\section{Spatial discretization via the finite element method}
\label{sec:spatdisc}
In order to use the aforementioned bifurcation tracking approaches on spatially continuous domains, it is necessary to obtain the discretized residual vector including the time dynamics, i.e. $\resvec$, from a spatially continuous residual functional $\resfunc$. Here, the finite element method is used for this step, but in principle, any other spatial discretization method can be applied. Particular focus is put on the complications that arise when a moving mesh is considered, i.e. the mesh coordinates are part of the unknowns. 

\subsection{Finite element method for a static mesh}
The finite element method provides a flexible and accessible approach to discretize arbitrary coupled equation in space. In particular, it can be applied to complicated geometries and moving meshes. In order to keep the equations brief, we will discuss it on the basis of a simple Poisson equation, although it does not show any bifurcations. Of course, it is textbook material, but it sets the basis to discuss all nonlinearities arising due to the moving mesh. 

The Poisson equation on a domain $\bulkdomain$ with boundary $\interfdomain=\interfdomain_\text{D}\cup\interfdomain_\text{N}$ reads
\begin{align}
\nabla^2 u&=h\quad\text{ on }\bulkdomain \nonumber\\
u&=u_\text{D}\quad \text{ on } \interfdomain_\text{D} \label{eq:poisson}\\
-\unitnormal\cdot \nabla u&=j\quad \text{ on } \interfdomain_\text{N} \nonumber \,,
\end{align}
i.e. with source term $h$, and Dirichlet and Neumann boundary conditions $u_\text{D}$ and $j$, respectively. The generic weak formulation reads
\begin{align}
\int_\bulkdomain \left( \nabla u\cdot\nabla \testfunc + h \testfunc\right){\rm d}\bulkdomain+ \int_{\interfdomain_\text{N}} j\testfunc\,{\rm d}\interfdomain=0\quad\text{ with }\quad u=u_\text{D}\quad \text{ on }\quad \Gamma_D \, ,
\end{align}
which has to hold for all appropriate choices of the test function $\testfunc$ that fulfill $\testfunc|_{\interfdomain_\text{D}}=0$.

The central idea of the finite element method is the choice of a discrete amount of basis functions for the test function $\testfunc$ and simultaneously also expand the unknown function $u$ in such a finite amount of basis functions. For the Galerkin approach, both sets of basis functions are the same, i.e. we introduce the shape functions $\shapefuncother^\othernode{l}$ for discrete values of $u$, i.e. for $\othernode{l}=1,\ldots,N_u^\text{dof}+N_u^\text{D}$. Here, the first $N_u^\text{dof}$ unknown degrees of freedom are located in the bulk of $\bulkdomain$, whereas the remaining $N_u^\text{D}$ are known values on the Dirichlet boundary $\interfdomain_\text{D}$. Next, $u$ is expanded as $u\approx u^\othernode{l}\shapefuncother^\othernode{l}$ (with summation over $\othernode{l}=1,\ldots,N_u^\text{dof}+N_u^\text{D}$) into discrete amplitudes $u^\othernode{l}$. On $\interfdomain_\text{D}$, the corresponding value $u^\othernode{l}$ is known from the Dirichlet condition $u_\text{D}$ and it is not considered as degree of freedom. Subsequently, one sets $\chi=\shapefuncother^\othernode{k}$ for $\othernode{k}=1,\ldots,N_u^\text{dof}$, i.e. only for the degrees of freedom, not for the Dirichlet values. Thereby, one obtains the discrete residual vector $\statresvec$ via
\begin{align}
\rescompo_\othernode{k}=\int_\bulkdomain \left( u^\othernode{l}\nabla \shapefuncother^\othernode{l}\cdot\nabla \shapefuncother^\othernode{k} + h \shapefuncother^\othernode{k}\right){\rm d}\bulkdomain+ \int_{\interfdomain_\text{N}} j\shapefuncother^\othernode{k}\,{\rm d}\interfdomain \label{eq:poissonres}
\end{align}
On a static mesh, it is trivial to obtain the Jacobian by differentiation with respect to $u_\othernode{l}$
\begin{align}
\jaccompo_{\othernode{k}\othernode{l}}=\int_\bulkdomain \nabla \shapefuncother^\othernode{l}\cdot\nabla \shapefuncother^\othernode{k}{\rm d}\bulkdomain \label{eq:poissonjac}
\end{align}
and the Hessian $\hesstens$ vanishes here due to the linearity of the Poisson equation.

\subsection{Calculating the spatial integrals}
To calculate the spatial integrals occurring e.g. in \eqref{eq:poissonres} and \eqref{eq:poissonjac}, one usually splits the integrals in a sum over all elements, separated in the discretized bulk domain $\bulkdomainh$ and the discretized Neumann boundary $\interfdomainh_\text{N}$,
\begin{align}
\rescompo_\othernode{k}=\sum_{\bulkelem\in\bulkdomainh} \rescompo_{\bulkdomainh,\othernode{k}}^\bulkelem+\sum_{\interfelem\in\interfdomainh_\text{N}}\rescompo_{\interfdomainh,\othernode{k}}^\interfelem
=\sum_{\bulkelem\in\bulkdomainh}\int_\bulkelem \left( u^\othernode{l}\nabla \shapefuncother^\othernode{l}\cdot\nabla \shapefuncother^\othernode{k} + h \shapefuncother^\othernode{k}\right){\rm d}\bulkdomain+ \sum_{\interfelem\in\interfdomainh_\text{N}}\int_{\interfelem} j\shapefuncother^\othernode{k}\,{\rm d}\interfdomain \label{eq:elemsep}
\end{align}
and likewise for the Jacobian $\jaccompo_{\othernode{k}\othernode{l}}$ and potentially the Hessian $\hesscompo_{\othernode{k}\othernode{l}\othernode{m}}$. The occurring integrals over the elements only have support in terms of the shape functions $\shapefuncother^\othernode{k}$ and $\shapefuncother^\othernode{l}$ if the corresponding degrees of freedom are parts of the current element $\bulkelem$ (or $\interfelem$ on the Neumann boundary $\interfdomain_\text{N}$). This allows the calculation of the elemental integrals by iterating the index $\othernode{k}$ (and $\othernode{l}$ and potentially $\othernode{m}$ for the Jacobian and Hessian, respectively) over the degrees of freedom associated with each element only.

\begin{figure}[t]
\centering
\includegraphics[width=0.75\textwidth]{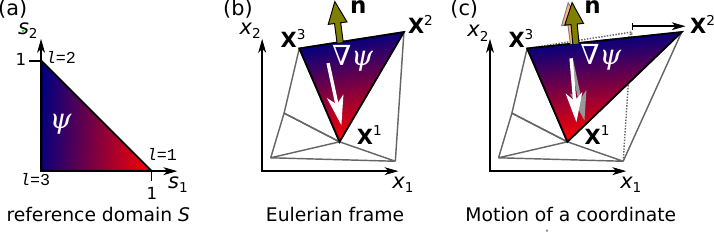}
\caption{(a) Spatial integrations are carried out element-wise, where the integrals are evaluated in the reference domain $S$. The value of the shape function $\shapefuncother$ corresponding to the node at $\vec{s}=(1,0)$ is indicated by the color gradient (blue: $\shapefuncother=0$, red: $\shapefuncother=1$). (b) By the Eulerian coordinates $\vec{X}^\posnode{l}$ of the element nodes, the integrals and derivatives are mapped into the Eulerian domain. (c) When a mesh node moves, the integral transformation, i.e. the area of the element, and also Eulerian gradients of shape functions $\nabla \shapefuncother^\othernode{l}$ and potentially normal vectors $\unitnormal$ at interfaces change accordingly. This gives plenty of additional contributions in the Jacobian and Hessian. }
\label{fig:scheme}
\end{figure}

The integration within each element is carried out in a fixed reference domain $S$, i.e. the local coordinate system $\vec{s}$ of the element (see figure~\ref{fig:scheme}). For the bulk contribution $\rescompo_{\bulkdomainh,\othernode{k}}^\bulkelem$, the integration is hence calculated by
\begin{align}
\rescompo_{\bulkdomainh,\othernode{k}}^\bulkelem=\int_{S} \left( u^\othernode{l}\nabla \shapefuncother^\othernode{l}\cdot\nabla \shapefuncother^\othernode{k} + h \shapefuncother^\othernode{k}\right)\sqrt{\det \mathbf{g}}\,{\rm d}^{n_\text{e}}s \,. \label{eq:localcoordint}
\end{align}
Here,  $n_\text{e}$ is the elemental dimension and the shape functions $\shapefuncother$ are directly evaluated in the reference domain $S$. The Eulerian derivatives $\nabla \shapefuncother$, however, must be transformed accordingly.
For that, the covariant metric tensor $\mathbf{g}$ ($n_\text{e}\times n_\text{e}$) of the coordinate transformation to the Eulerian coordinate $\vec{x}=\vec{x}(\vec{s})$ with the entries
\begin{align}
g_{\alpha\beta}=\vec{t}_\alpha\cdot\vec{t}_\beta\,.
\end{align}
is used. The covariant base vectors $\vec{t}_{\alpha}$ can be obtained by knowing that the Eulerian position $\vec{x}$ can also be expanded into shape functions $\shapefuncpos^\posnode{l}$, i.e. by
\begin{align}
\vec{x}=\vec{X}^\posnode{l}\shapefuncpos^\posnode{l}(\vec{s})
\end{align}
with the Eulerian coordinates $\vec{X}^\posnode{l}$ of the nodes and the shape functions of the position space $\shapefuncpos^\posnode{l}$. In general, the shape functions $\shapefuncpos^\posnode{l}$ of the Eulerian position $\vec{x}$ can be different from the shape functions $\shapefuncother^\othernode{l}$ of $u$, e.g. second order Lagrange basis functions for the position $\vec{x}$ and first order for $u$. In particular, also the range of $\posnode{l}$ and $\othernode{l}$ can be different. The covariant basis vectors can then be calculated from the nodal positions:
\begin{align}
\vec{t}_{\alpha}=\frac{\partial \vec{x}}{\partial s^\alpha}=\vec{X}^\posnode{l}\frac{\partial \shapefuncpos^\posnode{l}}{\partial s^\alpha}\,. \label{eq:covarbasevec}
\end{align}
Note that the dimension $n_\text{dim}$ of the Eulerian coordinate vector $\vec{x}$ can be higher than the element dimension $n_\text{e}$, e.g. on the interface elements $\interfelem$.

The Eulerian derivatives of the shape functions of $u$, i.e. $\nabla \shapefuncother$, in \eqref{eq:localcoordint} are transformed according to 
\begin{align}
\partial_{x_i}\shapefuncother^\othernode{l}=\left(\nabla \shapefuncother^\othernode{l}\right)_i=g^{\alpha\beta}t_{\alpha,i}\frac{\partial \shapefuncother^\othernode{l}}{\partial s^\beta}\,, \label{eq:spatialderivgab}
\end{align}
where $g^{\alpha\beta}$ is the contravariant metric tensor. On interface elements, this convention automatically yields the appropriate surface gradient operator. 

For interface elements $\interfelem$, also the facet normal $\unitnormal$ is calculated from the covariant base vectors \eqref{eq:covarbasevec}, where in some cases, the derivatives must actually be performed with respect to all nodal coordinates of the attached bulk element, e.g. for a one-dimensional interface element ($n_\text{e}=1$) attached to a two-dimensional bulk element ($n_\text{e}=2$) embedded in a three-dimensional Eulerian space ($n_\text{dim}=3$).

Eventually, the integration is carried out by a Gauss-Legendre quadrature on the reference domain $S$ including all these transformations.

\subsection{Additional nonlinearities stemming from a moving mesh}
If a moving mesh is considered, the Eulerian position $X_i^\posnode{l}$ of the mesh nodes are parts of the unknown vector $\dofvec$. For the Poisson equation \eqref{eq:poisson} defined on such a moving mesh, the vector of unknowns $\dofvec$, the residual vector $\statresvec$ and the Jacobian $\jacmat$ can hence be split into components of $u$ and the moving mesh coordinates $\vec{X}$:
\begin{align}
\dofvec=\begin{pmatrix} \dofvec_u \\ \dofvec_\vec{X} \end{pmatrix} \qquad \statresvec=\begin{pmatrix} \statresvec_u \\ \statresvec_\vec{X} \end{pmatrix} \qquad \jacmat=\begin{pmatrix} \jacmat_{uu} & \jacmat_{u\vec{X}} \\ \jacmat_{\vec{X}u} & \jacmat_{\vec{X}\vec{X}} \end{pmatrix}
\end{align}
The residual for the mesh motion is usually calculated on a fixed reference mesh, i.e. one describes the motion with respect to some Lagrangian coordinates, e.g. the initial mesh positions. Furthermore, we assume here for simplicity that the mesh motion is independent of $u$, i.e. $\jacmat_{\vec{X}u}=\mathbf{0}$. Any Eulerian integral or derivative contribution to the mesh position residuals $\statresvec_\vec{X}$ or any feedback of $u$ on $\vec{X}$, i.e. $\jacmat_{\vec{X}u}\neq\mathbf{0}$, must be treated equally as discussed in the following on the basis of the Jacobian block $\jacmat_{u\vec{X}}$.

While the Jacobian block $\jacmat_{uu}$ coincides with the one of a static mesh, plenty of additional terms arise in the block $\jacmat_{u\vec{X}}$. The derivative of the $\othernode{k}$-th elemental residual contribution \eqref{eq:localcoordint} of $u$ from the bulk with respect to the coordinate position $X^\posnode{l}_j$ reads
\begin{align}
\jaccompo_{u^\othernode{k}X^\posnode{l}_j}^\bulkelem=&\int_{S} \Big[ u^\othernode{l}\left(\partial_{X^\posnode{l}_j}\left(\nabla \shapefuncother^\othernode{l}\right)\cdot\nabla \shapefuncother^\othernode{k} +\nabla \shapefuncother^\othernode{l}\cdot\partial_{X^\posnode{l}_j}\left(\nabla \shapefuncother^\othernode{k}\right) \right)\sqrt{\det \mathbf{g}} \nonumber \\
&+ \left( u^\othernode{l}\nabla \shapefuncother^\othernode{l}\cdot\nabla \shapefuncother^\othernode{k} + h \shapefuncother^\othernode{k}\right)\partial_{X^\posnode{l}_j}\left(\sqrt{\det \mathbf{g}}\right)\Big]\,{\rm d}^{n_\text{e}}s \label{eq:jaccoorddiff}
\end{align}
To solve bifurcation tracking problems via Newton's method, also the Hessian is required. Entries from the Hessian block $\hesstens_{u\vec{X}u}$ can simply be calculated from \eqref{eq:jaccoorddiff} by deriving with respect to $u^\othernode{l}$. The elemental Hessian contribution from the bulk $\hesstens_{u\vec{X}\vec{X}}$ is obtained by deriving \eqref{eq:jaccoorddiff} again with respect to a potentially other nodal coordinate $X^\posnode{l'}_{j'}$. This gives an even longer expression than \eqref{eq:jaccoorddiff}, involving second order derivatives of the shape function gradients and the functional determinant $\sqrt{\det \mathbf{g}}$.
On interface elements $\interfelem$, the considered equation potentially also includes the normal vector $\unitnormal$, which first and second derivative with respect to the nodal positions is also required. In our implementation, we precalculate some quantities to efficiently calculate all these highly nonlinear derivatives up to the second order. These exact relations can be found in the supplementary information.

The example of a simple Poisson equation solved on a moving mesh illustrates the complexity that arises in the Jacobian and even more in the Hessian. Therefore, for more complicated problems, a sophisticated implementation to accurately calculate these derivatives automatically is required. This is discussed in the next section.

\section{Implementation}
\label{sec:implem}
All the complications discussed in the previous section can be circumvented if derivatives with respect to the moving mesh coordinates are evaluated numerically by finite differences, i.e. the entry $\jaccompo_{u^\othernode{k}X^\posnode{l}_j}$ of the Jacobian block $\jacmat_{u\vec{X}}$ is just calculated by
\begin{align}
\jaccompo_{u^\othernode{k}X^\posnode{l}_j}=\frac{\rescompo_{u^\othernode{k}}(\dofvec_u,\dofvec_{\vec{X}}+\epsilon \vec{e}^{\posnode{l}}_j)-\rescompo_{u^\othernode{k}}(\dofvec_u,\dofvec_{\vec{X}})}{\epsilon}
\end{align}
for some small value of $\epsilon$ which perturbs the nodal coordinate $X^\posnode{l}_j$ only, i.e. $\vec{e}^{\posnode{l}}_j$ is zero everywhere except at the equation index of $X^\posnode{l}_j$, where it is unity. For problems without bifurcation tracking, i.e. without the requirement of the Hessian, it works reasonably well. When the Hessian is required for bifurcation tracking problems, our numerical experiments have proven that finite differences of second order do not provide sufficient numerical accuracy (cf. table~\ref{tab:convrates} later). Newton's method for the augmented bifurcation tracking system usually does not converge, and if it does, it converges poorly. Also, the assembly of the augmented Jacobian is typically quite computationally expensive when using finite differences.

A promising way to overcome this lack of accuracy is automatic differentiation, but this requires the augmentation of all mathematical operations to account for dual numbers. While even third or higher order derivatives can be calculated quite efficiently and accurately up to machine precision by automatic differentiation, it can be complicated to apply this on the transformation from the local coordinate $\vec{s}$ to the Eulerian coordinate $\vec{x}$, which is rather hard-coded in many existing finite element frameworks. Furthermore, it is not trivial to reuse already calculated subexpressions that appear multiple times in the full Jacobian or Hessian during automatic differentiation. In general, the overhead due to automatic differentiation can increase the computational costs for assembly by a factor of up to two compared to hand-coded routines filling the Jacobian and/or Hessian \cite{Salinger2014}. 

In our implementation, we therefore opted for symbolical differentiation, i.e. indeed performing the steps discussed in the previous section, but entirely assisted and automatized by symbolic computer algebra. Only the residual must be defined in weak formulation and, subsequently, all symbolical derivatives up to second order, also with respect to the mesh coordinates, can be calculated. For performance reasons, all of these expressions are written in efficient C code which is automatically compiled and linked back into the running program. For the symbolic differentiation and the code generation, we use the efficient, accurate and flexible open-source framework \software{GiNaC} \cite{Bauer2002}. During the compilation, common subexpression elimination can speed up the process. Furthermore, particular quantities appearing multiple times in the residual can be explicitly marked by the user to be calculated and derived in beforehand. The performance of the generated code is on par with hand-coded implementations and typically, for multi-physics problems, even run up to twice as fast as the handwritten implementations of \software{oomph-lib}, as discussed in the supplementary material. Our approach of treating the entered equations symbolically, i.e. not by pure automatic differentiation, also has benefits in automatically deriving the corresponding forms for azimuthal symmetry breaking, as discussed later in section~\ref{sec:azibreak}.

As finite element framework, \software{oomph-lib} is employed \cite{Heil2006,Hazel2019}. It already offers several methods for bifurcation tracking and arclength continuation, monolithically treated moving meshes and support for multi-domain and multiphysics problems including spatial adaptivity, etc. The default way of calculating the Jacobian and Hessian, however, is by finite differences, unless the user explicitly codes the corresponding symbolical expressions by hand. Our automatic code generation fills exactly this gap, which is otherwise very cumbersome. \software{oomph-lib} also allows to access the low-level of the finite element method, i.e. the transformation from the local coordinate $\vec{s}$ to the Eulerian coordinate $\vec{x}$. Therefore, it provides the ideal framework for our purposes here.

Inspired by the framework \software{FEniCS} \cite{Logg2012}, we wrapped the core of \software{oomph-lib} into \software{python}, so that equations, problems and meshes can easily be assembled in \software{python}, but still use the full computational efficiency of the compiled \software{oomph-lib} core and the generated C code for the residual vector, the mass and Jacobian matrices, the Hessian and parameter derivatives of the former three. Arbitrary multi-physics problems can be formulated with a few lines of \software{python} code, where the entered residual form is directly converted to a \software{GiNaC} expression tree within our \software{C++} core. Rather arbitrary combinations of finite element spaces (including discontinuous Galerkin spaces) for scalar, vectorial and tensorial quantities can be used. Integral constraints, (fields of) Lagrange multipliers, potentially only defined at interfaces, and error estimators for mesh adaptivity can be incorporated directly. At shared interfaces between two multi-physics domains, the interface and bulk fields and gradients thereof can be accessed on both sides. Once the residual of the system is formulated that way, the monolithic Jacobian and Hessian can be quickly assembled on the basis of the generated C code. A more detailed overview of our framework and some example codes of our validation cases and are discussed in the supplementary information to illustrate the straightforward formulation of complicated problems.

\section{Validation}
\label{sec:validation}
To validate our implementation, we compare all types of bifurcation trackers with problems discussed in the literature in the following. Often, we use pseudo-arclength continuation in another parameter to obtain entire bifurcation diagrams, which usually only takes a few minutes with our framework.

\begin{table}
\scriptsize
\centering
\caption{Averaged assembly times for one Newton step (in seconds, on a single Intel i7-4790K thread/core) and number of Newton iterations (in brackets) for different methods the different bifurcation tracking examples discussed in the following. \textsl{dnc} means \textsl{does not converge}, i.e. Newton's method fails. (a) all derivatives symbolically calculated. (b) Hessian $\hesstens$ by finite differences from the symbolical Jacobian, all other derivatives symbolical. (c) Like (b), but all derivatives (also first order) with respect to the mesh coordinate by finite differences. (d) All derivatives by finite differences. Hopf tracking with pure finite differences are not available (\textsl{N/A}), since the filling of the mass matrix via finite differences is not implemented. For the different pitchfork cases, cf. section~\ref{sec:pitchtrack}.}
\begin{tabular}{|l|c|c|c|c|}
\cline{2-5}
\multicolumn{1}{c|}{}& (a) & (b) & (c) & (d)\\
\hline
fold (with Stokes flow)  & 0.85 (6) &  2.73 (6) &  4.63 (dnc) & 7.24 (dnc) \\
fold (with Young-Laplace equation)  & 0.77 (3) & 1.31 (4) & 1.64 (dnc) & 1.53 (dnc) \\
pitchfork (symmetric mesh) & 8.45 (2) & 14.20 (2) & 23.75 (dnc) & 28.67 (dnc) \\
pitchfork (nonsymm. mesh, dofwise symmetry) & 5.03 (dnc) & 9.31 (dnc) &  16.20 (dnc) & 19.82 (dnc) \\
pitchfork (nonsymm. mesh, weak symmetry) & 5.16 (16) & 11.35 (16) & 23.95 (dnc) & 31.52 (dnc) \\
Hopf & 5.33 (3) & 5.85 (3) & N/A & N/A  \\
\hline
\end{tabular}
\label{tab:convrates}
\end{table}

Before the discussion of the individual validation cases, however, the reader is referred to table~\ref{tab:convrates}. For all cases, we compared the assembly time of the augmented Jacobian $\augjacmat$ and the convergence of the Newton method for different assembly methods. Our symbolical method is always the fastest in assembly time and ensures good convergence for all considered cases. When calculating the second order derivatives in the Hessian by finite differences, but all first order derivatives symbolically, the convergence is mainly the same, but the assembly is slower. If any of the first order derivatives are calculated by finite differences, we could not achieve reliable convergence in any case. These results highlight the necessity of accurate derivatives in bifurcation tracking and the strengths of our symbolical code generation.

\subsection{Fold bifurcation}
\label{sec:foldbif}
To validated the fold bifurcation tracking on moving meshes, we consider a droplet hanging on the bottom of a vertical plate. Gravity will pull the droplet down and deform the droplet to deviate from a spherical cap shape according to the Young-Laplace equation, but at some critical volume $V$ or sufficiently strong gravitational force compared to the capillary force, the droplet will detach, i.e. the stationary hanging droplet solution will vanish in a fold bifurcation. 

In nondimensionalized quantities, we set the volume to unity, and use the Bond number $\Bo$ as bifurcation parameter. The Bond number can be calculated from the dimensional quantities as $\Bo=\rho g V^{2/3}/\sigma$, with the mass density $\rho$, the gravitational acceleration $g$ and the surface tension $\sigma$. As additional parameter, the nondimensionalized contact line radius $\rcontactline$, which is assumed to be fixed (pinned contact line) is introduced. Since the hanging solution corresponds to vanishing velocity, it is sufficient to solve the Stokes equations for the flow. The onset of detachment is also independent on the viscosity of the droplet, so we set it to unity. These assumptions obviously only hold for the bifurcation, not for the full detachment and pinch-off process, where inertia and viscous forces definitely play an important role \cite{Eggers1997}.

In total, we hence solve the following system in axisymmetric coordinates:
\begin{align}
\nabla\cdot\left[\nabla\vec{u}+(\nabla\vec{u})^T\right]&=\nabla p - \Bo \vec{e}_z\\
\nabla \cdot \vec{u}&=0\\
\left[\dot{\vec{X}}-\vec{u}\right]\cdot\unitnormal&=0 \label{eq:foldkinbc}\\
\left[\nabla\vec{u}+(\nabla\vec{u})^T\right]\cdot \unitnormal-p\unitnormal&=\kappa\unitnormal\\
 \int_{\text{drop}} 2\pi r {\rm d}r{\rm d}z&=1 \label{eq:foldvolconstr}
\end{align}
Here, $\kappa=-\nabla_S\cdot\unitnormal$ is the curvature and the volume constraint \eqref{eq:foldvolconstr} is enforced by the constant of the pressure nullspace with respect to an additive constant. The only time derivative appears is the motion of the mesh coordinate with the fluid velocity in normal direction, i.e. in the kinematic boundary condition \eqref{eq:foldkinbc}. The mesh dynamics can be chosen arbitrarily, as long as the mesh is following the physics of the fluid and does not influence the dynamics in a nonphysical way, i.e. despite entering via the curvature $\kappa$ in the dynamic boundary condition. Here, we just chose a Laplace-smoothed mesh. The weak formulation of this problem is available in the supplementary information. Also, the simple and convenient way to express this system within our code framework is illustrated within the supplementary information.

\begin{figure}[t]
\includegraphics[width=\textwidth]{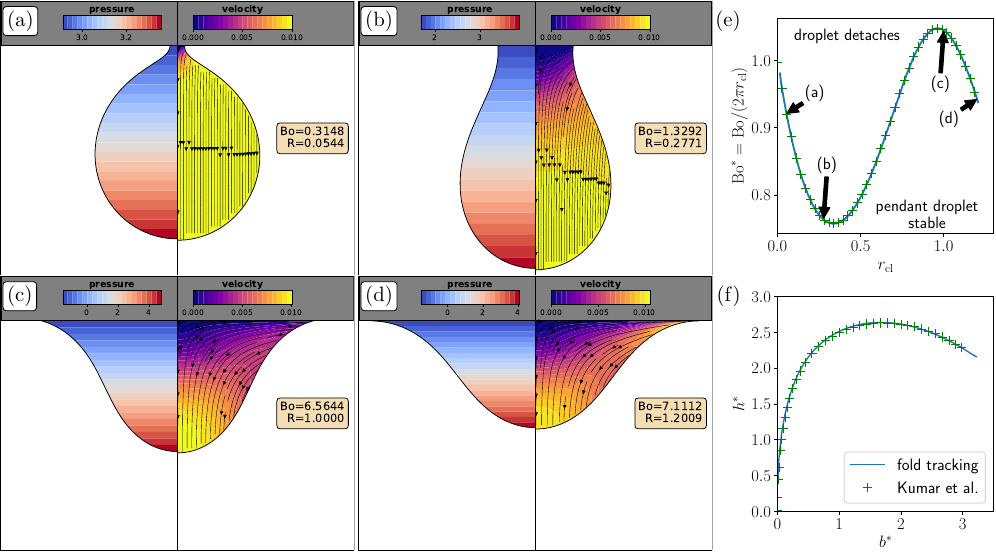}
\caption{Validation of the fold bifurcation tracking based on a hanging droplet detaching due to gravity. (a-d) snapshots of solutions on the bifurcation curve with critical solution (pressure, left) and critical eigendynamics (velocity and displacement, right) (e) Bifurcation curve expressed in $\Bo^*$ as function of the nondimensional contact radius $\rcontactline$. The markers stem for the fold bifurcation tracking of the Young-Laplace equation. (f) Validation with literature results extracted from Kumar et al. \cite{Kumar2020a,Kumar2020b}.}
\label{fig:foldvali}
\end{figure}

In figure~\ref{fig:foldvali}, the results of the fold bifurcation tracking are shown. The left sides of figure~\ref{fig:foldvali}(a-d) show some representative solutions at the critical Bond number at the given nondimensional contact line radius $\rcontactline$. The dynamics of the instability, extracted from the critical eigenfunction, are shown on the right half of each plot. While droplets with small $\rcontactline$ just start to fall down as spherical object, flow at near the contact line becomes more important for high values of $\rcontactline$. 

Figure~\ref{fig:foldvali}(e) shows a rescaled critical Bond number $\Bo^*=\Bo/(2\pi\rcontactline)$. This definition gives the ratio of gravitational force by the droplet mass to the capillary force acting on the circumference of the contact line. In the limit $\rcontactline\to 0$, it converges to unity as expected, i.e. the gravity by the entire droplet mass must be balanced by the capillary force at the contact line of the small connection with the wall. In total, however, it shows some nontrivial curve, which was determined by the fold bifurcation tracking combined with continuation in $\rcontactline$ in a few minutes. Occasionally, the mesh had to be reconstructed during this scan to prevent large mesh deformations. By interpolating the current critical solution and eigenfunction from the old mesh to the newly constructed mesh, the continuation can carry on after each mesh reconstruction. At $\rcontactline\approx 1.207$, the contact angle becomes zero before the critical Bond number can be reached. Here, the method cannot be continued due to the collapse of the element directly at the contact line. It is hence questionable whether a droplet with a pinned contact line $\rcontactline\gtrsim 1.207$ can detach at all. In reality, the contact line will depin or form a thin precursor film, which is not accounted for in this simple model.

Since the dynamics is entirely given by the interface, the bulk of the droplet is actually not required to obtain the critical curve. We also solved the Young-Laplace equation directly on a line mesh embedded in a two-dimensional axisymmetric coordinate system. The equation reads
\begin{align}
-\nabla_S\cdot \unitnormal+\Bo z&=p_\text{ref} \label{eq:foldyl} \\
\frac{2\pi}{3}\int \unitnormal\cdot\vec{x}\:r{\rm d}l+1&=0
\end{align}
The second equation enforces a unit volume via the reference pressure $p_\text{ref}$. The Young-Laplace equation \eqref{eq:foldyl} is solved on moving mesh coordinates in normal direction, whereas for the tangential degrees of freedom, equidistant positioning of the nodes along the arclength is enforced. The radial mesh coordinate at the axis is fixed to zero and the axial mesh coordinate is fixed to zero at the contact line, where also the radial position is prescribed to be $\rcontactline$. The same fold bifurcation algorithm as before applied on this equation gives the markers in figure~\ref{fig:foldvali}(e), i.e. perfect agreement, but considerably faster due to the lack of bulk equations. However, the method including the bulk domain is flexible to be extended, e.g. to include the effect of thermal Marangoni convection, when a nonuniform temperature profile is induced in the droplet by heating of the top wall. Then, continuation can be performed in the thermal Marangoni number, which is however not within the scope of this article.

Finally, to validate our results, the critical droplet shape was expressed in new parameters, namely $b^*=\rcontactline\sqrt{\Bo}$ and $h^*=h\sqrt{\Bo}$ (with the nondimensional height of the droplet $h$). The comparison with the data extracted from Refs. \cite{Kumar2020a,Kumar2020b} also gives nice agreement, as shown in figure~\ref{fig:foldvali}(f).

\subsection{Pitchfork and Hopf bifurcation}
\label{sec:vali:pitchhopf}
\begin{figure}[t]
\includegraphics[width=\textwidth]{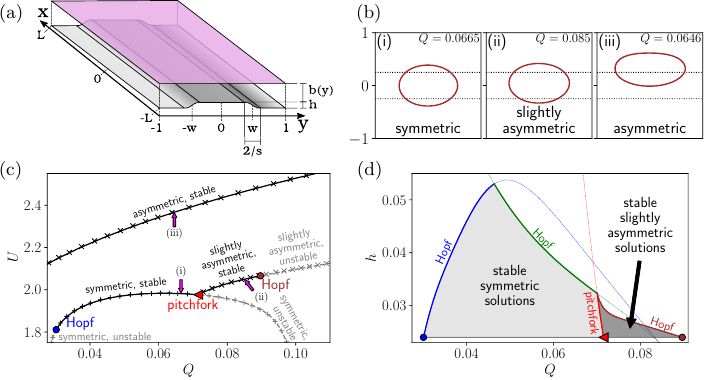}
\caption{Validation of the pitchfork and Hopf bifurcation tracking based on a bubble in a Hele-Shaw cell with a centered constriction. (a) Sketch of the problem with the relevant geometric parameters. (b) Representative types of stationary solutions of the bubble. (c) Validation of our results (lines) with the results extracted from Keeler et al. \cite{Keeler2019} (points) for $h=0.024$. Arrows indicate the solutions depicted in (b). (d) Bifurcation tracking in the height of the bump $h$.  }
\label{fig:pfhopfvali}
\end{figure}

For the pitchfork and Hopf bifurcation on moving meshes, an ideal validation case can be found in the articles of Thompson et al. on the solutions of a bubble propagating through a Hele-Shaw cell with a transversal depth perturbation \cite{Thompson2014,FrancoGomez2017,Keeler2019,Thompson2021,Gaillard2021}. As depicted in Fig.~\ref{fig:pfhopfvali}(a), a narrow Hele-Shaw cell that has a constriction in the center is considered. When bubbles are advected through this channel in $x$-direction, they can either move to the left or right side ($y$-direction), where the channel is higher. However, in some parameter ranges, stable symmetric solutions are also possible, where the bubble propagates along the center line ($y=0$) of the bump. As shown by Keeler et at. \cite{Keeler2019}, this symmetric centered state can loose the stability in either Hopf or pitchfork bifurcations to an asymmetric state.

Due to the shallow cell, it is sufficient to consider lubrication theory (or potential flow, Darcy's law) to describe the problem, i.e. in the liquid surrounding the bubble, a Laplace equation for the pressure field $p(x,y)$ is solved. Due to the varying nondimensional local height of the channel $b(y)$, the equation actually reads
\begin{align}
\nabla\cdot\left(b^3 \nabla p\right)=0\,,
\end{align}
and the nondimensional velocity field follows to be $\vec{u}=-U \vec{e}_x-b^2\nabla p$. Here, $U$ is the bubble velocity in lab frame, i.e. the velocity $\vec{u}$ is expressed in the frame co-moving with the bubble. The bubble velocity $U$ is part of the unknowns and follows from the constraint that enforces the averaged $x$-coordinate of the bubble this co-moving frame to zero. Likewise, the pressure inside bubble $p_\text{B}$ is assumed to be homogeneous, i.e. inviscid and inertia-free gas. This pressure $p_\text{B}$ corresponds to the volume constraint
 \begin{equation}
 \int_\text{bubble} b\,\mathrm{d}x\mathrm{d}y=V
 \end{equation}
 for some given volume $V$ and enters the dynamic boundary condition
 \begin{equation}
p_\text{B}-p=\frac{1}{3\alpha Q}\left(\frac{1}{b}+\frac{\kappa}{\alpha}\right)\,,
 \end{equation}
where $Q$ is a non-dimensional flow rate or capillary number with respect to mean through flow velocity of the channel. The aspect ratio, i.e. the channel width in $y$-direction divided by the plate distance, is denoted by $\alpha$, whereas $\kappa$ denotes the $x$-$y$-projected curvature of the bubble and $1/b$ accounts for the second curvature \cite{Keeler2019}. As exterior boundary conditions, $\partial_y p|_{y=\pm 1}=0$, $\partial_x p|_{x=-L}=-1$ and $p|_{x=L}=0$ are imposed. The nondimensional constricted height is given by a smoothed double step with a bump height $h$, sharpness $s$ and half-width $w$, i.e.
\begin{equation}
b(y)=1-\frac{h}{2}\left[\tanh\left(s(y+w)\right)-\tanh\left(s(y-w)\right)\right]\,.
\end{equation}
Lastly, the kinematic boundary condition demands $b^2\nabla p\cdot\unitnormal=-U \unitnormalcompo_x-\dot{\vec{X}}\cdot\unitnormal$.

In accordance with Keeler et al. \cite{Keeler2019}, we set $h=0.024$, $w=0.25$, $L=4$, $\alpha=40$, $s=40$ and fix the projected bubble volume to $V=\pi r^2$ with $r=0.46$. Our implementation yields the same bistable region as described by Keeler et al. \cite{Keeler2019}, where the three types of stationary solutions are represented in Fig.~\ref{fig:pfhopfvali}(b). The stability diagram also agrees perfectly with the data extracted from their work (cf. Fig.~\ref{fig:pfhopfvali}(c)), where lines are our results and dots are the results of Keeler et al. The bifurcation points in Fig.~\ref{fig:pfhopfvali}(c) were obtained by bifurcation tracking. Once a bifurcation point is located, one can obtain entire bifurcation diagrams in minutes by continuation, as e.g. shown exemplary for an increase in the bump height $h$ in Fig.~\ref{fig:pfhopfvali}(d). With increasing height $h$, the bistable region shrinks, the pitchfork and the Hopf bifurcation of the slightly asymmetric branch merge and are dominated by another Hopf bifurcation branch. Eventually, at $h\approx 0.053$, symmetric solutions cannot be found anymore for any flow rate $Q$. A detailed analysis of the entire parameter space is of course not within the scope of this article. However, with our method of dynamic code generation, it is also straightforward to formulate generalizations of this problem, e.g. considering Stokes flow with a Brinkman term instead of potential flow. This would allow to imposed correct tangential boundary conditions, e.g. no-slip boundaries at the side walls of the channel.

\section{Azimuthal stability analysis of axisymmetric base states}
\label{sec:azibreak}
Our method to symbolically evaluate the Hessian has proven to work well in the previous validation section. In principle, it can also be applied to full three-dimensional problems, but the monolithic treatment will result in huge augmented Jacobian matrices which are cumbersome to invert for Newton's method. Direct solvers will run out of memory and finding suitable preconditioners for an iterative solver is complicated for the augmented problem.

If a bifurcation occurs for a base solution with particular symmetry, one can use this symmetry to reduce the problem size, e.g. from full three-dimensional to axisymmetric cylindrical coordinates, but still allow for instabilities that break this symmetry. Here, we focus on axisymmetric problems that lose the azimuthal symmetry in a bifurcation, but the same method can also be applied if the base state is e.g. invariant in the third Cartesian dimension.

\subsection{Method outlined on a static mesh}
\label{sec:staticmeshazimethod}
For simplicity, we first develop the method of static meshes and subsequently generalize it for moving meshes. This numerical approach has been e.g. applied by Yim et al. pancake vortices in stratified liquids \cite{Yim2016}, but also e.g. to numerically predict the self-propulsion of Leidenfrost droplets \cite{Yim2022}. Recently, also the symmetry breaking due to thermal Marangoni flow, both in non-volatile droplets on heated or cooled substrates \cite{Babor2023} and in rotating annular pools \cite{Liu2021}, has been investigated on the basis of the following method.

Axisymmetric problems can best be described in axisymmetric cylindrical coordinates $(r,z)$, i.e. an axisymmetric stationary solution $\doffuncaxi$ is just a function of $(r,z)$. However, to also allow for perturbations breaking this symmetry, the residual formulation $\resfunc$ must also account for derivatives with respect to the azimuthal coordinate $\aziangle$, e.g. $\nabla f=(\partial_r f,r^{-1}\partial_\aziangle f,\partial_z f)$ for any scalar function $f$. If this is ensured, the linear evolution of a azimuthally perturbed state
\begin{align}
\doffunc^\text{pert}(r,\aziangle,z,t)=\doffuncaxi(r,z)+\epsilon e^{\lambda_m t}e^{im\aziangle}\doffuncazi(r,z)+\text{c.c.} \label{eq:azipertsubst}
\end{align}
can be considered. The problem is, however, still formulated on a two-dimensional mesh, i.e. using the reduction by the symmetry, whereas the azimuthal dynamics are entirely given by the mode $e^{im\aziangle}$. The goal is to find the complex-valued eigenfunction $\doffuncazi$ and the corresponding eigenvalue $\lambda_m$ for a given azimuthal mode with integer-valued mode number $m$. In linear order in the parameter $\epsilon\ll1$, these modes are independent and upon linerization one obtains
\begin{align}
\resfunc(\doffunc^\text{pert})\approx\epsilon e^{\lambda_m t}\left(\lambda_m\massop+\jacop\right)e^{im\aziangle}\doffuncazi(r,z)+\text{c.c.} = 0\,. \label{eq:contiazieigen}
\end{align}
Here, the operators $\massop$ and $\jacop$ are the continuous analogues of the discretized mass matrix and the Jacobian, i.e. the G\^{a}teaux-derivatives of $\resfunc$ with respect to $\partial_t\doffunc$ and $\doffunc$ evaluated at $0$ and $\doffuncaxi$, respectively. Both operators may contain $\aziangle$-derivatives, i.e. they are acting on the product $e^{im\aziangle}\doffuncazi(r,z)$. Obviously, for \eqref{eq:contiazieigen} to hold after spatial discretization, i.e. $(\doffunc^0,\mathcal{V}^m,\mathcal{M},\mathcal{J}) \to(\dofvec^0,\eigenvec^m,\massmat^m,\jacmat^m)$, one has to solve the generalized discretized eigenproblem
\begin{align}
\lambda_m \massmat^m{\eigenvec}^m=-\jacmat^m{\eigenvec}^m \label{eq:azieigen}
\end{align}
which differs from \eqref{eq:geneigen} by having an $m$-dependent mass and Jacobian matrix. Furthermore, $\massmat^m$ and $\jacmat^m$ are in general complex-valued now, since odd derivatives with respect to $\aziangle$ induce imaginary contributions.

Since our framework is based on a symbolical treatment of the entered residual formation, it is capable to derive the expressions necessary to assemble $\massmat^m$ and $\jacmat^m$ automatically from arbitrary weak formulations of the residual functional $\resfunc$. Before any derivatives are applied in the entered form of $\resfunc$ or any spatial discretization is performed, first all scalar fields $f$ and vector fields $\vec{u}$ in $\doffunc$ are expanded according to \eqref{eq:contiazieigen}, i.e.
\begin{align}
    f(r,\aziangle,z,t)&\to f^0(r,z,t)+\epsilon f^m(r,z,t)e^{im\aziangle}\\ \nonumber
    \vec{u}(r,\aziangle,z,t)&\to\vec{u}^0(r,z,t)+\epsilon \vec{u}^m(r,z,t)e^{im\aziangle}\,.
\end{align}
The fields $f^0$ and $\vec{u}^0$ are entries of the axisymmetric stationary solution function $\doffunc^0$, whereas $f^m$ and $\vec{u}^m$ are part of the eigenmode $\mathcal{V}^m$. Although $f^0$ and $\vec{u}^0$ stationary solutions, i.e. are not explicitly time-dependent, and the time-dependence of $f^m$ and $\vec{u}^m$ is later replaced by $e^{\lambda_m t}$, an arbitrary time dependency is considered here in order to automatically generate the mass matrix entries.
The corresponding test functions $g$ and $\vec{v}$ are replaced by
\begin{align}
    g(r,z) \to  g(r,z)e^{-im\aziangle}\,\qquad \vec{v}(r,z) \to \vec{v}(r,z)e^{-im\aziangle}
\end{align}
The vectorial fields and test functions are additionally augmented by an additional component $\aziangle$-direction during this step.
Potential global degrees of freedom, e.g. a Lagrange multiplier that enforces a volume constraint or removes e.g. the nullspace of some field, are not expanded, i.e. only the base mode corresponding to $m=0$ is kept.

After plugging in, the original axisymmetric residual formulation can be recovered by setting $\epsilon=m=0$, whereas the in general complex-valued auxiliary azimuthal residual ${\resfunc}^m$ is obtained by the first order Taylor coefficient in $\epsilon$.

For the example, an axisymmetric diffusion equation for a scalar field $f$, i.e. $\partial_t f=\nabla^2 f$ (without Neumann terms and with test function $g$) yields the following axisymmetric residual $\mathcal{R}^{*,0}$ and auxiliary azimuthal residual ${\resfunc}^m$ after applying this method:
\begin{align}
    \resfunc(\doffunc)&=2\pi  \int_\Omega \left(\partial_t fg+\partial_r f \partial_r g + \partial_z f \partial_z g \right)   r\: {\rm d}r{\rm d}z \\
    \resfuncazi(\doffunc^0,\mathcal{V}^m)&=2\pi  \int_\Omega \left(\partial_t f^m g+\partial_r f^m \partial_r g + \partial_z f^m \partial_z g +\frac{m^2}{r^2}fg \right)   r\: {\rm d}r{\rm d}z\,. \label{eq:diffueqazires}
\end{align}
Nonlinear terms in the original residual would give additional couplings between the axisymmetric stationary solution $\doffunc^0$ and the azimuthal perturbation $\mathcal{V}^m$, i.e. between $f^0$ and $f^m$ in this example case.

After this, spatial discretization can be performed, e.g. expansions in shape functions and, by our method, performant C code is generated for both types of residuals and the corresponding mass matrices, Jacobian matrices, potential parameter derivatives and Hessians. 
The matrices for the azimuthal eigenproblem \eqref{eq:azieigen} are then obtained from the discretized auxiliary residual ${\resvec}^m$ via
\begin{align}
    \massmat^m=\frac{\partial {\resvec}^m}{\partial \dot{\eigenvec}^{m}}\qquad\text{and}\qquad \jacmat^m=\frac{\partial {\resvec}^m}{\partial \eigenvec^{m}}
    \label{eq:massandjacm}
\end{align}
evaluated at $\statdofvec$ and with $\dot\statdofvec=\dot\eigenvec^m=0$. Since the azimuthal perturbation $\mathcal{V}^m$ enters only linearly in ${\resfunc}^m$, both of these matrices are independent on ${\eigenvec}^{m}$, but may depend on the axisymmetric stationary solution $\statdofvec$ due to nonlinearities.

An axisymmetric stationary solution $\statdofvec$ is obtained as before, i.e. by Newton's method with the generated code corresponding to $\resfunc$. This solution can then be investigated for stability and bifurcations as previously to assess for stability with respect to axisymmetric perturbations ($m=0$). But the axisymmetric solution $\statdofvec$ can additionally be checked for axisymmetry-breaking instabilities ($m\neq 0$) by solving the azimuthal eigenproblem \eqref{eq:azieigen} for different values of $m$. For physically reasonable problems, the range of $m$ is limited since, for high values of $m$, stabilizing terms will dominate and yield only eigenvalues with negative real parts, as e.g. due to the term $m^2/r^2$ in \eqref{eq:diffueqazires} stemming from azimuthal diffusion. A single line in the driver code entered by the user automatically invokes the code generation for all required matrices, so that arbitrary problems can be investigated for axial symmetry breaking easily.

\subsection{Boundary conditions for the eigenvalue problem}\label{subsec:azibc}
Particular care must be taken with the boundary conditions at the axis of symmetry $r=0$. For the axisymmetric base state, scalar fields have to fulfill $\partial_r f^0=0$, likewise the axial component of vector fields fulfill $\partial_r u^0_z =0$, whereas the radial and azimuthal component have to vanish, i.e. $u^0_r=0$ and $u^0_\aziangle=0$. For the eigenvector corresponding to the azimuthal perturbation, the boundary conditions at the axis of symmetry depend on $m$. For $m=0$, we solve the conventional eigenvalue problem \eqref{eq:geneigen}, with the same boundary conditions, but for $|m|=1$, the boundary conditions at $r=0$ must be changed to $f^m=u^m_z=0$ and $\partial_r u^m_r=\partial_r u^m_\aziangle=0$, since the basis of the vector components exactly rotates with the mode $m=1$ and for $|m|\geq 2$, all components have to vanish, i.e. $f^m=u^m_r=u^m_\aziangle=u^m_z=0$ \cite{Batchelor1962,Yim2016,Yim2022}.

Integral constraints associated with a global degree of freedom, e.g. a Lagrange multiplier enforcing some volume or some spatial average of a field, must be deactivated for $m\neq 0$ as well. Due to the rotation by $e^{im\aziangle}$, an azimuthal perturbation with $m\neq 0$ always has a vanishing contribution to these constraints when considering the full three-dimensional problem, i.e. the corresponding entry in the eigenvector $\eigenvec^m$ has to be removed.

Depending on the choice of $m$, our framework automatically takes care of imposing the correct boundary conditions at $r=0$ and toggling potential integral constraints. This is achieved by modifying the assembled matrices in the eigenvalue problem \eqref{eq:azieigen} accordingly, depending on the current value of $m$.

\subsection{Bifurcation tracking for azimuthal symmetry breaking}\label{sec:azimuthal_bifurcation_tracking}

We are interested in generalizing the bifurcation tracking approaches discussed in section~\ref{sec:biftrack} to azimuthal instabilities, i.e. in finding the critical parameter $\critparam$ for which the axisymmetric stationary solution breaks its azimuthal symmetry. The following bifurcation tracking method for azimuthal symmetry breaking allows to use two-dimensional discretizations to solve this problem. It consists in solving the eigenproblem \eqref{eq:azieigen} with the goal of finding the critical parameter $\critparam$ for which the eigenvalue $\lambda^m=\lambda^m_\mathrm{r}+i\lambda^m_\mathrm{i}$ crosses the imaginary axis, i.e. $\lambda^m_\mathrm{r}(\critparam)=0$. In order to do so, the discretized residual vector from the axisymmetric base state, $\resvec$, is augmented with the eigenproblem \eqref{eq:azieigen} and with a constraint to avoid the trivial eigensolution $\eigenvec^m=\eigenvec^m_\mathrm{r}+i\eigenvec^m_\mathrm{i}=\vec{0}$. The mass and Jacobian matrices, $\massmat^m=\massmat^m_\mathrm{r}+i\massmat^m_\mathrm{i}$ and $\jacmat^m=\jacmat^m_\mathrm{r}+i\jacmat^m_\mathrm{i}$, corresponding to the $m$-dependent azimuthal auxiliary residual function $\resfuncazi$ (cf. \eqref{eq:massandjacm}) are assembled to the augmented residuals ${\augresvec}^{,m}$ and augmented Jacobian ${\augjacmat}^{,m}$:
\begin{gather}
    {\augdofvec}^{,m}=\begin{pmatrix} \dofvec \\ \eigenvec^m_\mathrm{r} \\ \eigenvec^m_\mathrm{i} \\ p \\ \eigenval_\mathrm{i} \end{pmatrix}\qquad 
    {\augresvec}^{,m}=\begin{pmatrix} \statresvec \\ \jacmat^m_\mathrm{r} \eigenvec^m_\mathrm{r} - \jacmat^m_i \eigenvec^m_\mathrm{i} - \eigenval_\mathrm{i} (\massmat^m_\mathrm{r} \eigenvec^m_\mathrm{i} + \massmat^m_\mathrm{i} \eigenvec^m_\mathrm{r}) \\ \jacmat^m_\mathrm{i} \eigenvec^m_\mathrm{r} + \jacmat^m_\mathrm{r} \eigenvec^m_\mathrm{i} + \eigenval_\mathrm{i} (\massmat^m_\mathrm{r} \eigenvec^m_\mathrm{r} - \massmat^m_\mathrm{i} \eigenvec^m_\mathrm{i}) \\ \eigenvec^m_\mathrm{r}\cdot\vec{C}-1 \\ \eigenvec^m_\mathrm{i}\cdot\vec{C}\end{pmatrix}  \nonumber \\ 
    {\augjacmat}^{,m}=\begin{pmatrix} \jacmat  & \mathbf{0} & \mathbf{0} & \partial_p\statresvec & \mathbf{0} \\ \partial_\dofvec ({\augresvec}^{,m}_2) & \jacmat^m_\mathrm{r} - \eigenval_\mathrm{i} \massmat^m_\mathrm{i}  & -\jacmat^m_\mathrm{i} - \eigenval_\mathrm{i} \massmat^m_\mathrm{r} &  \partial_p ({\augresvec}^{,m}_2) & -\massmat^m_\mathrm{r} \eigenvec^m_\mathrm{i} - \massmat^m_\mathrm{i} \eigenvec^m_\mathrm{r} \\ \partial_\dofvec ({\augresvec}^{,m}_3) & \jacmat^m_\mathrm{i} + \eigenval_\mathrm{i} \massmat^m_\mathrm{r} & \jacmat^m_\mathrm{r} - \eigenval_\mathrm{i} \massmat^m_\mathrm{i}  & \partial_p ({\augresvec}^{,m}_3) & \massmat^m_\mathrm{r} \eigenvec^m_\mathrm{r} - \massmat^m_\mathrm{i} \eigenvec^m_\mathrm{i} \\ 0 & \vec{C} & 0 & 0 & 0 \\ 0 & 0 & \vec{C} & 0 & 0 \end{pmatrix} \,.
\end{gather}

For brevity, the abbreviation ${\augresvec}^{,m}_i$ in the augmented Jacobian ${\augjacmat}^{,m}$ corresponds to the $i^\mathrm{th}$ block row of the augmented residual vector ${\augresvec}^{,m}$. The augmented system is more involved than the one required for e.g. Hopf bifurcations due to the fact that the mass and Jacobian matrices $\massmat^m$ and $\jacmat^m$ are complex-valued. The constraint vector $\vec{C}$ is usually set to the initial guess of the eigenvector $\vec{V}$. If the initial guess is sufficiently close to the bifurcation, the Newton method \eqref{eq:newton} can be used to solve the augmented system, yielding the stationary axisymmetric solution ${\dofvec}_0$, the azimuthal perturbation eigenvector $\eigenvec^m$, the imaginary part of the corresponding eigenvalue ${\eigenval}_\mathrm{i}$ and the critical parameter $\critparam$.
Since the azimuthal eigenproblem has complex-valued matrices, it is not necessary to distinguish between the different bifurcation types, e.g. fold, pitchfork or Hopf bifurcation.

\subsection{Validation on a static mesh}

As a validation example of the bifurcation tracking for azimuthal symmetry breaking, we consider the Rayleigh-Bénard convection problem in a cylindrical container, as analyzed by Borónska and Tuckerman \cite{Boronska2006}. A fluid confined within a cylindrical boundary and heated from below, as shown in Fig. \ref{fig:RBschematics}(a), changes from a motionless conductive state to a flow with convective rolls, due to the action of gravity, when a critical temperature difference $\Delta T_c$ between the bottom and top layers is reached. We refer to the Rayleigh number $Ra$ to characterize the temperature difference and to a critical $Ra_c$ for the flow to settle in. The characteristics of these rolls depend on $Ra$, the Prandtl number $Pr$ and on the cylinder's aspect ratio $\Gamma = r/h$,  where $r$ is the radius of the cylinder and $h$ is its height. The rolls can be either axisymmetric or non-axisymmetric, depending on the critical azimuthal mode $m$ of which corresponding eigenvector settles the instability that generates the rolls. For instance, if for a certain $Ra$ there is an instability solely for $m=2$, then the stationary solution will have four convective rolls in the cylinder, as depicted on the streamlines of the velocity in Fig. \ref{fig:RBschematics}(b). 

The used model equations here are the nondimensionalized Boussinesq equations, which correspond to the Navier-Stokes equations with the Boussinesq approximation for the density, and the advection diffusion equation for temperature $T$:

\begin{figure}[t]
\centering
    \includegraphics[width=0.7\textwidth]{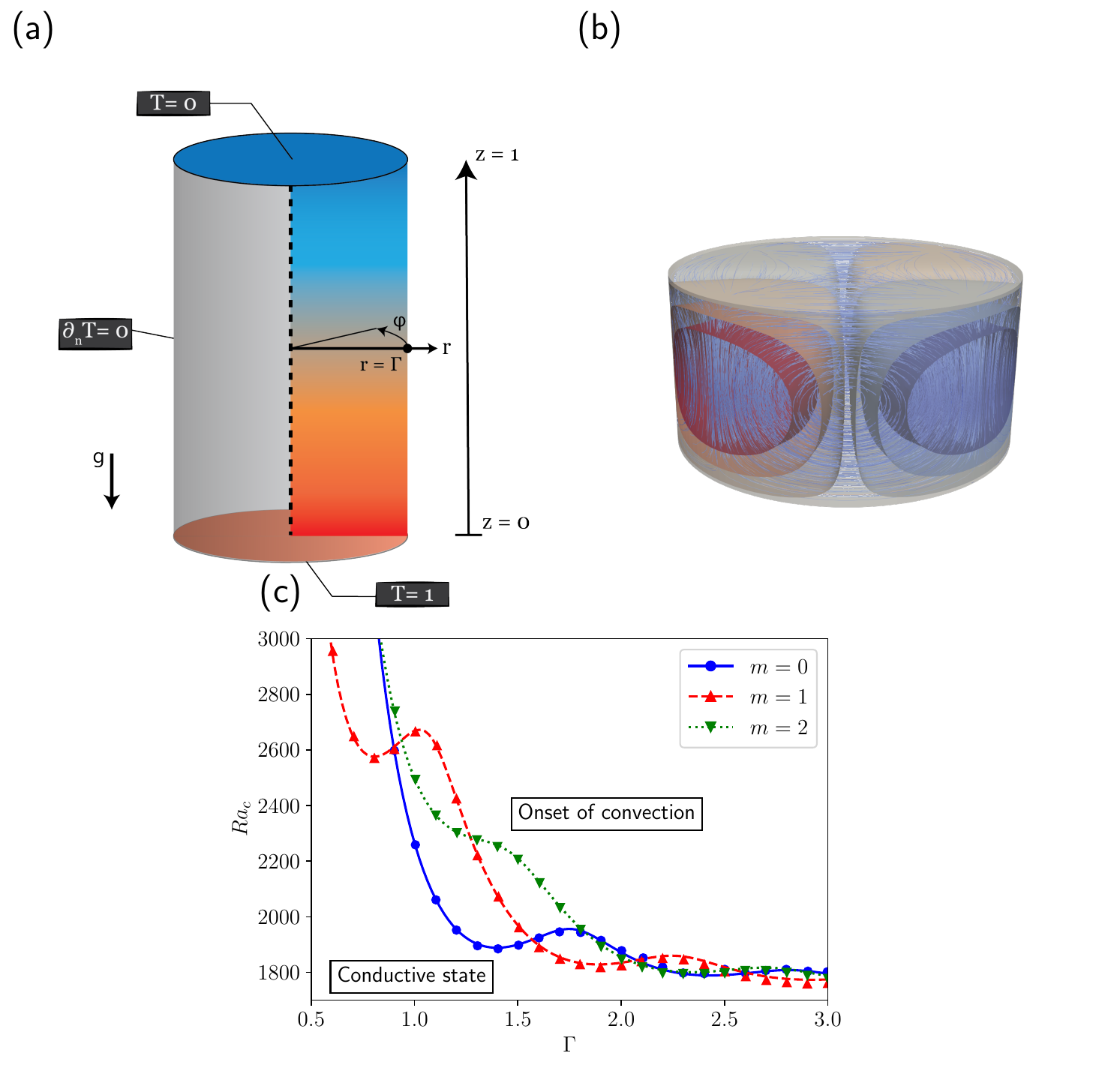}
    \caption{Rayleigh-B\'enard in a cylindrical container. (a) Sketch of the geometry and temperature's boundary conditions. (b) Eigenfunction of temperature field and streamlines of velocity for stationary solution at $Ra_c$, with aspect ratio $\Gamma = 1$ and azimuthal mode $m=2$. (c) Comparison of the critical Rayleigh number $Ra_c$ for the onset of convection for different azimuthal modes $m=\{0,1,2\}$ between our results (lines) and the ones in \cite{Boronska2006} (points).}
    \label{fig:RBschematics}
\end{figure}

\begin{align}
\nabla \cdot \vec{u} &= 0 \label{eq:RBdivu}\\
\partial_t \vec{u} + \vec{u} \cdot \nabla \vec{u} &= - Ra Pr\nabla p + Pr \nabla^2 \vec{u} + Pr Ra T \vec{e}_z \label{eq:RBconvection1}\\
\partial_t T + \vec{u} \cdot \nabla T &= \nabla^2 T \label{eq:RBconvection2} \, .
\end{align}

\noindent Furthermore, in coherence with the boundary conditions used in \cite{Boronska2006}, we impose no-slip boundary conditions on the velocity field at the cylinder's walls. The temperature is fixed to $T_0=1$ at $z=0$ and to $0$ at $z=1$, while at the sidewalls of the cylinder are adiabatic, i.e. $\nabla T \cdot \unitnormal = 0$ at $r = \Gamma$, as indicated on Fig. \ref{fig:RBschematics}(a). In accordance to the analysis of \cite{Boronska2006}, we consider only aspect ratios $\Gamma \sim 1$.
The instability of the conductive state is independent of $Pr$, so $Pr=1$ in our simulations. We use a two-dimensional rectangular domain to investigate the $Ra_c$ for which the flow becomes unstable for each $m=\{0,1,2\}$. Starting by fixing $m$ to one of the prescribed values, we find a $Ra$ value which is, for an initial $\Gamma$, close to the bifurcation, to supply a good initial guess for the critical eigenvector. Then, we augment our system through the method explained on the section \ref{sec:azimuthal_bifurcation_tracking} and we solve it through the Newton method, using the initial guess of the critical eigenvector at the first iteration. Lastly, we use arclength continuation on $\Gamma$ to obtain the bifurcation curve corresponding to each $m$. Note that we scale the $r$-coordinate with the $\Gamma$ parameter, thus not requiring a constant remeshing in order to obtain these curves. The results are depicted in Fig. \ref{fig:RBschematics}(c), showing a good agreement between our results (lines) and the ones in \cite{Boronska2006} (points). 

It should be noted that the viscous term in \eqref{eq:RBconvection1} leads to a $1/r^2$ singularity for $m=1$ and $r\to 0$ in the azimuthal perturbation, which is analytically not integrable \cite{Gelfgat1999,Babor2023}. A mathematical elegant treatment is the usage of the continuity equation to replace the viscous term \cite{Gelfgat1999,Babor2023}, but apparently, due to the enforcing of the continuity equation and the fact that the Gauss-Legendre quadrature never evaluates at $r=0$, our method is also numerically stable. In fact, without using the reformulation of the viscous term for $m=1$, we still could perfectly reproduce the results of Ref. \cite{Babor2023}, as we show in the supplementary information.

\subsection{Method generalized for moving meshes}
In the following, we combine the described methods of bifurcation tracking on moving meshes and the investigation of the azimuthal stability of axisymmetric base states. Typical problems that undergo an azimuthal instability by changing the shape are e.g. capillary bridges beyond the Steiner limit \cite{Lowry1995} or bucking of elastic tubes due to capillary effects, as e.g. analyzed by Hazel \& Heil \cite{Hazel2005}. 

Of course, these problems can be implemented straight-forward on a three-dimensional Cartesian mesh, but for the azimuthal linear stability, it is sufficient to reduce the dynamics around the axisymmetric stationary solution to a two-dimensional mesh, i.e. evaluated at $\aziangle=0$, and apply the normal mode expansion in terms of $\exp(im\aziangle)$ on it, as described in section \ref{sec:staticmeshazimethod}. However, for this step, it is crucial to also add perturbations to the mesh coordinates. In a general three-dimensional cylindrical coordinate system, the position vector of the axisymmetric base state is given by
\begin{align}
\posvecaxi=r^0\unitvec_r(\aziangle)+z^0\unitvec_z
\end{align}
and the corresponding perturbed position vector reads $\posvecaxi+\epsilon \posvecazi e^{im\aziangle}$ with
\begin{align}
\posvecazi=r^m\unitvec_r+r^0 \aziangle^m\unitvec_\aziangle +z^m\unitvec_z\,.
\end{align}
If the azimuthal mesh position does not have a physical meaning, i.e. it is just used for parametrization, $\aziangle^m$ can be set to zero, but in general cases, e.g. for the torsion of an elastic body, $\aziangle^m$ must be kept as unknown part of the perturbation, which then must be solved for in the azimuthal eigenvalue problem. As before, during the spatial discretization, both $\posvecaxi$ and $\posvecazi$ are expanded in terms of the position shape functions $\shapefuncpos^l$:
\begin{align}
r^0&=X^{0\posnode{l}}_r \shapefuncpos^\posnode{l}\,,  &z^0&=X^{0\posnode{l}}_z \shapefuncpos^\posnode{l} &\\
r^m&=X^{m\posnode{l}}_r \shapefuncpos^\posnode{l}\,,  &z^m&=X^{m\posnode{l}}_z \shapefuncpos^\posnode{l}  \,, \qquad &\aziangle^m=X^{m\posnode{l}}_\aziangle \shapefuncpos^\posnode{l} 
\end{align}
i.e. in the discrete nodal positions and the complex-valued corresponding azimuthal eigenvector, respectively.

Likewise, the normal $\unitnormal$ changes with the perturbation. During the perturbation, the normal of the axisymmetric base state  
\begin{align}
\normalvecaxi=\normalcompoaxi_r\unitvec_r+\normalcompoaxi_z\unitvec_z
\end{align}
must be replaced by a corresponding linearized perturbed normal, $\normalvecaxi+\epsilon \normalvecazi e^{im\aziangle}$, where the change of the normal (in general not of unit length) due to the perturbation in linear order in $\epsilon$ is given by
\begin{align}
\normalvecazi=\left(\frac{\partial n_r^0}{\partial X^{0\posnode{l}}_j} X^{m\posnode{l}}_j\right)\unitvec_r + \left(\normalcompoaxi_r \aziangle^m -\frac{im}{r^0}\normalvecaxi\cdot\posvecazi \right)\unitvec_\aziangle + \left(\frac{\partial n_z^0}{\partial X^{0\posnode{l}}_j} X^{m\posnode{l}}_j\right)\unitvec_z
\end{align}
Additionally, the differential operators must be extended accordingly. Without azimuthally perturbed mesh coordinates, we calculate e.g. the divergence of a vector field $\vec{v}=\vec{v}^0+\epsilon e^{im\aziangle}\vec{v}^m$ by
\begin{equation}
\nabla \cdot \vec{v}=g^{\alpha\beta}t_{\alpha,r}\frac{\partial v_r}{\partial s^\beta}+g^{\alpha\beta}t_{\alpha,z}\frac{\partial v_z}{\partial s^\beta}+\frac{v_r}{r}+\frac{1}{r}\partial_\aziangle v_\aziangle\,,
\end{equation}
i.e. the metric tensor just accounts for the mapping from the local element coordinates $s_\beta$ to Cartesian Eulerian coordinates (cf. \eqref{eq:spatialderivgab}), while additional terms from the cylindrical coordinate system are added afterwards. Due to the presence of the azimuthal perturbed coordinates, its first order expansion in $\epsilon$ reads
\begin{align}
\nabla \cdot \vec{v}&=g^{\alpha\beta}t_{\alpha,r}\frac{\partial v_r^0}{\partial s^\beta}+g^{\alpha\beta}t_{\alpha,z}\frac{\partial v_z^0}{\partial s^\beta}+\frac{v_r^0}{r^0} \nonumber \\
&+\epsilon e^{im\aziangle}\left(g^{\alpha\beta}t_{\alpha,r}\frac{\partial v_r^m}{\partial s^\beta}+g^{\alpha\beta}t_{\alpha,z}\frac{\partial v_z^m}{\partial s^\beta}+\frac{v_r^m}{r^0}+\frac{1}{r^0}\partial_\aziangle v_\aziangle^m\right) \\ 
&+\epsilon e^{im\aziangle}\left( \left[D^{\posnode{l}\beta}_{rr}X_r^{m\posnode{l}}+D^{\posnode{l}\beta}_{rz}X_z^{m\posnode{l}}\right] \frac{\partial v_r^0}{\partial s^\beta}+\left[D^{\posnode{l}\beta}_{zr}X_r^{m\posnode{l}}+D^{\posnode{l}\beta}_{zz}X_z^{m\posnode{l}}\right]\frac{\partial v_z^0}{\partial s^\beta}-\frac{r^m}{(r^0)^2}v_r^0  \right) \nonumber \,.
\end{align}
While the first two lines are already present in the azimuthal stability analysis without a moving mesh, the third line considers the effect of the linear azimuthal perturbation of the mesh coordinates. Here, we use the abbreviation
\begin{equation}
D^{\posnode{l}\beta}_{ij}=\left.\partial_{X_j^{\posnode{l}}}\left( g^{\alpha\beta}t_{\alpha,i}\right)\right|_{\vec{X}^0}\,,
\end{equation}
i.e. the derivatives of the transformation terms with respect to the mesh coordinates. While they appear to be cumbersome, they are in fact already calculated in beforehand for the symbolical Jacobian of the moving mesh. Relations for $D^{\posnode{l}\beta}_{ij}$ are available in the supplementary material.

All these additional expansions are performed automatically within our framework, if a moving mesh is considered. After the expansion, the $m$-dependent azimuthal auxiliary residual function $\resfuncazi$ is again obtained by the first order in $\epsilon$, from which the azimuthal eigenvalue problem matrices are subsequently derived.

\subsection{Validation for moving meshes}
Due to the complexity, it is hard to find a good validation case for azimuthal shape instabilities in literature. Here, we consider a capillary surface as before in section~\ref{sec:foldbif}, but now in the configuration of a liquid bridge between two cylindrical plates with pinned contact lines and in absence of gravity. Long bridges undergo a Rayleigh-Plateau instability, whereas liquid bridges with high volumes (compared to the volume of a cylinder between the two plates) and small plate distances show asymmetric states. The entire stability dynamics has been already investigated in quite some detail \cite{Lowry1995,Bostwick2015}, so that in particular the transition to asymmetric states provides a suitable validation case of our azimuthal bifurcation tracking method. The bifurcation curve to non-axisymmetric bridges is known \cite{Slobozhanin1997}. It happens exactly when the liquid volume exceeds the threshold so that the capillary surface becomes tangent to the contact line, which is the limit when the simplifying Steiner symmetrization is not possible anymore \cite{Lowry1995,Slobozhanin1997,Bostwick2015}.

\begin{figure}[t]
\centering
    \includegraphics[width=0.6\textwidth]{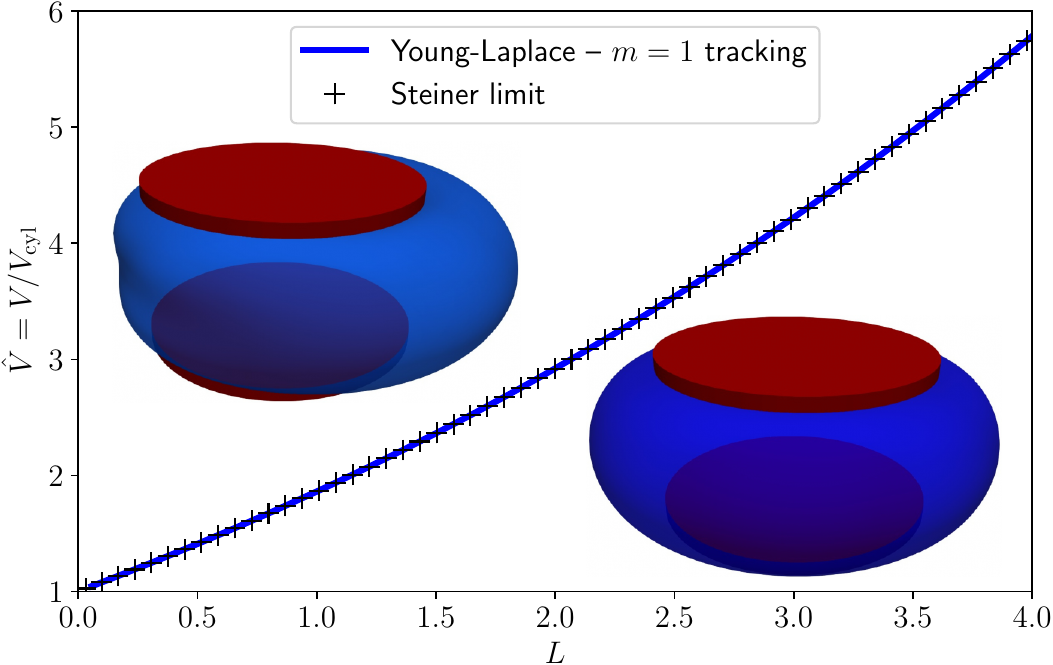}
    \caption{Azimuthal bifurcation tracking on a moving mesh on the basis of a liquid bridge. If the normalized liquid volume $\hat{V}$ trapped between the two cylindrical plates with distance $L$ exceeds a threshold, an $m=1$ instability of the shape is observed. Our azimuthal bifurcation tracking method, applied on a moving mesh, perfectly recovers the theoretical Steiner limit. The insets show the axisymmetric base state (lower right) and the base state plus the critical $m=1$ eigenfunction (upper left) at $L=1$ and $\hat{V}\approx 1.8603$.}
    \label{fig:liqbridge}
\end{figure}

The axisymmetric base states of a liquid bridge can easily be solved by the Young-Laplace equation, where the capillary surface is represented by a revolved line mesh with coordinates $(r,z)$ in axisymmetric cylindrical coordinates and the corresponding outward pointing normal $\unitnormal$:
\begin{align}
\nabla_S\cdot \unitnormal&=p\,. \label{eq:ylbridge}
\end{align}
The end points of the surface line are fixed at the nondimensional coordinates $(1,0)$ and $(1,L)$, where $L$ is the plate distance. The pressure (or curvature) $p$ is constant along the surface. $p$ is either given as a parameter or, alternatively, constitutes a Lagrange multiplier enforcing a prescribed normalized volume $\hat{V}$ by virtue of
\begin{align}
\frac{2\pi}{3}\int \unitnormal\cdot\vec{x}\:r{\rm d}l-\pi L\left(\hat{V}-\frac{1}{3}\right)&=0\,.
\end{align}
The normalized volume $\hat{V}$ is the ratio of the actual volume of the liquid divided by the volume of a cylinder with height $L$ and radius $1$.

We can solve this system easily by our moving mesh capabilities and continue the solution e.g. in $V$ and $p$ or $\hat{V}$, respectively. However, this only gives access to  axisymmetric $m=0$ instabilities. By applying the described azimuthal bifurcation tracking, however, the limit of asymmetric states can be found as well without any further changes in the code. The automatic code generation takes care of expanding all quantities and operators, like the normal $\unitnormal$ and its surface divergence $\nabla_S\cdot\unitnormal$, to the corresponding azimuthally perturbed variants, from what the azimuthal eigenvalue problem can be assembled. We can jump on the symmetry-breaking $m=1$ bifurcation by automatically adjusting $p$ correspondingly and subsequently perform pseudo-arclength continuation in the bridge length $L$ to obtain the curve shown in Fig.~\ref{fig:liqbridge}.

\section{Conclusion}
We have developed and validated a numerical method that allows stability analysis and bifurcation tracking on arbitrary multi-physics problems. Particular complications induced by problems with moving domains, i.e. on moving meshes, are tackled by exact symbolical derivatives of the entered system residual up to the second order, including the derivatives with respect to the moving mesh coordinates. On the basis of the symbolically obtained forms for the residual, the Jacobian and mass matrix, parameter derivatives thereof, as well as the Hessian, efficient C code is automatically generated, which ensures high performance. Due to the numerical exact treatment, our approach does not only outperform the trivially implemented finite-difference approach, but also ensures good convergence of Newton's methods applied to the augmented bifurcation tracking systems. For the latter, we have proven that finite-difference methods generically fail due to the inexact calculation of the derivatives, in particular on moving mesh problems.

Our method has been successfully validated on the basis of versatile literature results. By combining the bifurcation tracking with continuation, entire phase diagrams in the parameter space can be obtained within minutes. The definition of the entire equation system, including the geometry, parameters and potentially additional equations, usually takes only $\sim 100$ lines of easily readable \software{python} code, even for nontrivial multi-physics problems.

For complicated three-dimensional settings, however, the approach can still be rather expensive. As a workaround, our method has been generalized to automatically investigate azimuthal symmetry breaking instabilities of axisymmetric stationary solutions. Thereby, the symmetry of the base state is fully utilized, i.e. again allowing for quick calculations on an axisymmetric mesh in two spatial dimensions only, but yet extracting the full three-dimensional instabilities.

With this method, it is envisioned to investigate a plethora of bifurcations in fluid dynamics which are hardly accessible by analytical methods.  Due to the moving mesh capability, our framework will e.g. easily allow to find the Hopf bifurcation for the onset of bouncing of a droplet in a stratified liquid due to an interplay of Marangoni and Rayleigh forces, as e.g. reported in \cite{Li2019,Li2021,Li2022,Meijer2023,Herrada2023}. With the azimuthal symmetry-breaking analysis, also the onset of nonaxisymmetric flow fields in evaporating droplets due to solutal \cite{Diddens2017} or thermal \cite{Sefiane2008} Marangoni flow can be analyzed. Likewise, the motion of a Leidenfrost droplet due to a $m{=}1$-instability can be investigated at a finite capillary number and including the entire gas phase dynamics, and thereby generalizing the analysis of Yim et al. \cite{Yim2022}. Due to the moving mesh capability, also the onset of motion of an inverse Leidenfrost droplet levitating on a bath  \cite{Gauthier2019} can be obtained. The method furthermore could be applied to the autochemotactic motion of active droplets \cite{Maass2016,Michelin2023}, toroidal liquid films \cite{Edwards2021}, and to a plethora of more interesting scenarios. 

While the bifurcation tracking can find the location and general type of the bifurcation, weakly nonlinear dynamics is not accessible, i.e. in particular it cannot reveal whether a bifurcation is super- or subcritical. Normally, transient simulations in the vicinity of the bifurcation can bring clarity, but for our azimuthal symmetry approach, only the linear dynamics is available. Here, the automatic code generation of our symbolic framework could easily derive a weakly nonlinear generalization of \eqref{eq:azipertsubst}, i.e. including quadratic of cubic order in $\epsilon$, including the nonlinear coupling between different, nonlinearly excited, azimuthal modes. For moving meshes, however, this might be too complicated due to the nonlinear changes of e.g. the normals. Also, the developed bifurcation trackers could be generalized to find codimension-two bifurcations (cf. e.g. \cite{Kuznetsov1998,Bindel2014}) or the stability of limit cycles in the future.

\section*{Acknowledgement}
This work was supported by an Industrial Partnership Programme of the Netherlands Organisation for Scientific Research (NWO) \& High Tech Systems and Materials (HTSM), co-financed by Canon Production Printing Netherlands B.V., University of Twente, and Eindhoven University of Technology. The authors thank Dr. Alice Thompson, Dr. Jack Keeler and Dr. Lukas Babor for providing additional information, data and source code. 


\clearpage
\pagebreak

\begin{center}
\textbf{\large \textbf{Supplementary Information}\\Bifurcation tracking on moving meshes and with consideration of azimuthal symmetry breaking instabilities}\ \\
{\large Christian Diddens, Duarte Rocha}
\end{center}
\setcounter{equation}{0}
\setcounter{figure}{0}
\setcounter{table}{0}
\setcounter{page}{1}
\setcounter{section}{0}
\renewcommand{\thesection}{S-\Roman{section}}
\makeatletter
\renewcommand{\theequation}{S\arabic{equation}}
\renewcommand{\thefigure}{S\arabic{figure}}

\clearpage
\pagebreak

\section{First and second order derivatives with respect to the nodal mesh coordinates}
\label{si:sec:meshderivs}
When an Eulerian mesh node position $X^\posnode{l}_j$ is moved, the Eulerian derivatives of any field expanded in shape functions change accordingly, as elaborated in the main article. The Eulerian derivative of a shape function $\shapefuncother^\othernode{l}$ are evaluated by
\begin{align}
\partial_{x_i}\shapefuncother^\othernode{l}=g^{\alpha\beta}t_{\alpha,i}\frac{\partial \shapefuncother^\othernode{l}}{\partial s^\beta}\,,
\end{align}
The first and second order differentiation of this quantity with respect to the nodal coordinates $X^\posnode{l}_j$ (and $X^\posnode{l'}_{j'}$) yields
\begin{align}
\partial_{X^\posnode{l}_j}\left(\partial_{x_i}\shapefuncother^\othernode{l}\right)&=D^{\posnode{l}\beta}_{ij}\frac{\partial \shapefuncother^\othernode{l}}{\partial s^\beta}\\
\partial_{X^\posnode{l'}_{j'}}\partial_{X^\posnode{l}_j}\left(\partial_{x_i}\shapefuncother^\othernode{l}\right)&=E^{\posnode{l}\posnode{l'}\beta}_{ijj'}\frac{\partial \shapefuncother^\othernode{l}}{\partial s^\beta}
\end{align}
Here, the transformations $D$ and $E$ are calculated only once for each Gauss-Legendre integration point
\begin{align}
D^{\posnode{l}\beta}_{ij}&=\partial_{X^\posnode{l}_j}\left(g^{\alpha\beta} t_{\alpha,i}\right)= \contracoorddiff^{\posnode{l}\alpha\beta}_{j}t_{\alpha,i}+ \delta_{ij}g^{\alpha\beta} \partial_{s^\alpha}\shapefuncpos^\posnode{l} \\
E^{\posnode{l}\posnode{l'}\beta}_{ijj'}=&\partial_{X^{\posnode{l'}}_{j'}}\partial_{X^\posnode{l}_j}\left(g^{\alpha\beta} t_{\alpha,i}\right)=\partial_{X^{l'}_{j'}}D^{\posnode{l}\beta}_{ij} \nonumber\\
=&-D^{\posnode{l'}\gamma}_{ij'}\covarcoorddiff^{\posnode{l}}_{\gamma\delta j}g^{\delta\beta}-\delta_{jj'}g^{\alpha\gamma}\left(\partial_{s^\gamma}\shapefuncpos^\posnode{l}\partial_{s^\delta}\shapefuncpos^{\posnode{l'}}+\partial_{s^\gamma}\shapefuncpos^{\posnode{l'}}\partial_{s^\delta}\shapefuncpos^{\posnode{l}}\right)g^{\delta\beta} t_{\alpha,i} \nonumber \\&-g^{\alpha\gamma}\covarcoorddiff^{\posnode{l}}_{\gamma\delta j}\contracoorddiff^{\posnode{l'}\delta\beta}_{j'} t_{\alpha,i}+ \delta_{ij}\contracoorddiff^{\posnode{l'}\alpha\beta}_{j'}\partial_{s^\alpha}\shapefuncpos^\posnode{l}    
\end{align}
with the derivatives of the co- and contravariant metric tensor with respect to the nodal coordinates
\begin{align}
\covarcoorddiff^{\posnode{l}}_{\gamma\delta j}&=\partial_{X^\posnode{l}_j}g_{\gamma\delta}=\left(\partial_{s^\gamma}\shapefuncpos^\posnode{l}\right)t_{\delta,j}+\left(\partial_{s^\delta}\shapefuncpos^l\right)t_{\gamma,j}\\
\contracoorddiff^{\posnode{l}\alpha\beta}_{j}&=\partial_{X^\posnode{l}_j}g^{\alpha\beta}=-g^{\alpha\gamma}\covarcoorddiff^{\posnode{l}}_{\gamma\delta j}g^{\delta\beta} 
\end{align}
Of course, the symmetry of $E^{\posnode{l}\posnode{l'}\beta}_{ijj'}=E^{\posnode{l'}\posnode{l}\beta}_{ij'j}$ can be used for additional performance.

Likewise the functional determinant $\sqrt{\det\mathbf{g}}$, which appears when evaluating the Eulerian spatial integral in the elemental reference domain $S$, has contributions when derived with respect to moving nodal mesh coordinates, i.e.
\begin{align}
\partial_{X^\posnode{l}_j}\sqrt{\det\mathbf{g}}&=\sqrt{\det\mathbf{g}} A^\posnode{l}_j \\
\partial_{X^\posnode{l'}_{j'}}\partial_{X^\posnode{l}_j}\sqrt{\det\mathbf{g}}&=\sqrt{\det\mathbf{g}} B^{\posnode{l}\posnode{l'}}_{jj'}
\end{align}
with the following factors obtained by Jacobi's formula
\begin{align}
A^\posnode{l}_j&= g^{\alpha\beta}t_{\alpha,j}\frac{\partial \shapefuncpos^\posnode{l}}{\partial s^\beta}\\
B^{\posnode{l}\posnode{l'}}_{jj'}&= A^\posnode{l}_jA^\posnode{l'}_{j'}+D^{\posnode{l'}\beta}_{jj'}\frac{\partial \shapefuncpos^\posnode{l}}{\partial s^\beta}\,.
\end{align}


Interface elements will also contain additional contributions associated to the derivation of the normal vector with respect to the moving nodal mesh coordinates. Of course, the calculation of the normal depends on the nodal dimension considered. For instance, the unit normal $\unitnormal = \frac{\vec{n}}{||\vec{n}||}$ to one-dimensional interface elements associated with two-dimensional bulk elements can be embedded in a three-dimensional space, which makes its calculation more complex when compared with the unit normal to e.g. a two-dimensional interface element in a three-dimensional space. In the former case, it is considered as covariant vectors the tangent to the surface in the direction of the intrinsic surface coordinates, $\vec{t}_\alpha$, and the tangent in the direction of a bulk local coordinate that varies away from the interface (interior direction $\gamma$), $\vec{t}_\gamma$. The vectors are calculated by taking the derivative of the position with respect to the face coordinate $\vec{t}_\alpha = \partial_{s^\alpha} \vec{x}$, and the interior direction $\vec{t}_\gamma = \partial_{s^\gamma} \vec{x}$, respectively. By taking the cross product $\vec{t}_\alpha \times \vec{t}_\gamma$, one obtains the three-dimensional normal to the element, $\vec{n}_{3D}$. The normal to the interface element will be given by the cross product $\vec{n}_{3D} \times \vec{t}_\alpha$. This triple cross product can be handled applying the triple product expansion:

\begin{gather}
    \vec{n} = \vec{t}_\alpha \times \vec{t}_\gamma \times \vec{t}_\alpha = (\vec{t}_\alpha \cdot \vec{t}_\gamma) \vec{t}_\alpha - (\vec{t}_\alpha \cdot \vec{t}_\alpha) \vec{t}_\gamma \, .
\end{gather}

\noindent For simplicity, we write solely the derivation of the first order differentiation of $i$-component of the unit normal $\unitnormal$ with respect to the nodal moving coordinates $X^\posnode{l}_j$:

\begin{align}
    \partial_{X^\posnode{l}_j} \Big(\frac{n_i}{||\vec{n}||} \Big)&=\frac{\partial_{X^\posnode{l}_j} n_i}{||\vec{n}||} + \frac{n_i}{||\vec{n}||^3} \Big(n_p  \partial_{X^\posnode{l}_j} n_p\Big) = \frac{1}{||\vec{n}||} \Big( K^\posnode{l}_{ij} + \frac{n_p}{||\vec{n}||^2} K^\posnode{l}_{pj} \Big) \, ,
\end{align}

\noindent where the transformation $K^\posnode{l}_{ij}$ is given by:

\begin{align}
    K^\posnode{l}_{ij} = \partial_{X^\posnode{l}_j} n_i
    &= (\delta_{jk} \partial_{s^\alpha} \shapefuncpos^\posnode{l} t_{\gamma,k})t_{\alpha,i} + (\delta_{jk} t_{\alpha,k} \partial_{s^\gamma} \shapefuncpos^\posnode{l}) t_{\alpha , i} + (t_{\alpha , k} t_{\gamma , k}) \delta_{ij} \partial_{s^\alpha} \shapefuncpos^\posnode{l} \nonumber \\ &- 2(\delta_{jk} \partial_{s^\alpha} \shapefuncpos^\posnode{l} t_{\alpha , k}) t_{\gamma , k} - (t_{\alpha , k} \delta_{ij} t_{\alpha , i}) \partial_{s^\gamma} \shapefuncpos^\posnode{l} \, .
\end{align}

For discontinuous Galerkin methods, usually an estimator for the typical element size $h$ is required, e.g. the circumradius of the element. Of course, this also change with the mesh coordinates and similar relations can be calculated and used.

\section{General outline of our code framework}
\begin{figure}[t]
\includegraphics[width=\textwidth]{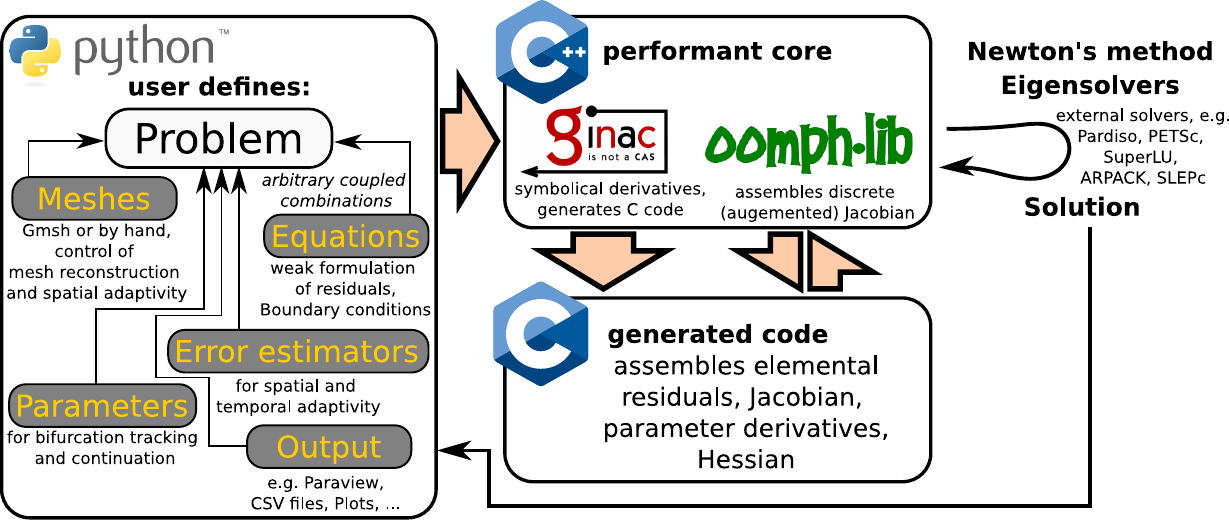}
\caption{Our code framework allows easy and flexible combination of meshes, equations, parameters and further settings directly in \software{python}. The calculations are done by a performant \software{C++} core, which first generates just-in-time compiled code from the entered equations. Also the corresponding code is generated for the Jacobian and Hessian, where \software{GiNaC} is used to perform the required symbolic differentiation. After compilation of the generated codes, \software{oomph-lib} monolithically assembles the (augmented) Jacobian and solves it via external linear algebra solvers or eigensolvers. This approach allows a fast and flexible setup by still having the maximum performance.}
\label{fig:codeframework}
\end{figure}

An overview of our code framework is schematically depicted in Fig. \ref{fig:codeframework}. Equations can easily be defined in \software{python}, where just the symbolical definition of the weak residual form is required. Equations can be defined in a coordinate-system agnostic way, i.e. independently of whether e.g. an Cartesian or a cylindrical coordinate system is eventually used. Also, arbitrary combinations of finite element spaces, including discontinuous Galerkin spaces, can be combined, e.g. for Taylor-Hood elements.

Meshes are also directly defined in \software{python}, which can be either done entirely manually, i.e. adding nodes and elements by hand, or invoke third-party meshing tools as e.g. \software{Gmsh} \cite{Geuzaine2009} to construct a mesh from a sketched geometry. For curved boundary, \software{oomph-lib}'s \cite{Heil2006}  macro-elements are automatically constructed to accurately represent these boundaries. 

Meshes and the equation system are combined in a problem class. Arbitrary combinations of equations can be merged, accounting for all the mutual couplings between the equations, e.g a Navier-Stokes equation and an advection-diffusion equation for a Rayleigh-Bénard system. Likewise, boundary conditions or initial conditions can be applied. The equation system is then merged with the meshes based on the names of the domains and interfaces. Thereby, the required Eulerian dimension and the dimension of the elements is available for the equations. The choice of the coordinate system can be set at problem level, but also at equation level or even for individual terms in the residuals within the equation classes. Thereby, spatial differential and integral operations can be carried out correctly. At equation level, interface equations applied on shared mutual interfaces between two domains can access the bulk fields and gradients thereof of both sides, as well as fields defined on the interface itself.

Once this step is done, a C code is generated for each domain and interface. This C code is made for a performant assembly of the elemental residuals. Likewise, code is generated for the Jacobian and potentially the Hessian, where the required symbolical differentiation of the entered weak residual formulation is obtained by \software{GiNaC}. This also includes the derivatives with respect to the mesh coordinates as discussed in section~\ref{si:sec:meshderivs}. Code to fill the parameter derivatives of the residual and the Jacobian is also generated. Likewise, code for initial conditions, Dirichlet boundary conditions and spatial error estimators for spatial adaptivity are written.

Subsequently, these C codes are compiled and loaded back into the code automatically. Whenever a transient or stationary solution -- or the solution of an eigenvalue problem -- is demanded, \software{oomph-lib} \cite{Heil2006} can handle the assembly of the required residual vectors, mass matrices and monolithic (augmented) Jacobian matrices. To that end, a reduced version of \software{oomph-lib} has been incorporated into our core and the element classes of \software{oomph-lib} were augmented to call the dynamically generated C code and generalized to account for the arbitrary finite element space combinations that may appear in the defined equations on each domain.

Our code framework hence combines the simple definition of problems and arbitrary equations and the subsequent code generation from \software{FEniCS} \cite{Logg2012}, while the object-oriented approach and the monolithic assembly of arbitrary multi-physics problems on multiple and potentially moving domains is taken over from the design idea of \software{oomph-lib}.

\section{Performance compared to other frameworks}
We compared the performance of the generated equation code to example cases from \software{oomph-lib} \cite{Heil2006}, mainly for simple transient or stationary solutions problems with or without moving meshes. 

Results of the assembly time comparison are shown in table~\ref{si:tab:speedcompare}. For a moving mesh case, we considered the example case \pyth{elastic_single_layer_interface} of \software{oomph-lib}, i.e. the relaxation of a perturbed fluid interface to flat equilibrium with gravity. The number of quadrilateral Crouzeix-Raviart elements for the Navier-Stokes equation and a pseudo-elastic mesh for the mesh motion was increased to $300 \times 300$, yielding $1\,350\,600$ degrees of freedom. For a simple test case on a static mesh, we used the two-dimensional Poisson equation example \pyth{two_d_poisson} of \software{oomph-lib} with $1000 \times 1000$ quadrilateral second-order elements, i.e. $4\times 10^6$ degrees of freedom.

\begin{table}
\scriptsize\centering
\begin{tabular}{|c|c|}
\multicolumn{2}{c}{{\large\pyth{elastic_single_layer_interface}}} \\ 
\hline
Method & Avg. assembly of $(\vec{R},\mathbf{J})$ [s] \\
\hline
\textsl{our code}, Jacobian fully symbolically derived automatically & 6.2 \\
\textsl{our code}, symbolical Jacobian derived automatically, mesh coordinate derivatives by FD & 10.2 \\
\software{oomph-lib}, symbolical Jacobian by hand, mesh coordinate derivatives by FD & 16.7 \\
\textsl{our code}, all derivatives by FD & 23.4 \\
\hline
\end{tabular}
\ \\ \ \\
\begin{tabular}{|c|c|}
\multicolumn{2}{c}{{\large\pyth{two_d_poisson}}} \\ 
\hline
Method & Avg. assembly of $(\vec{R},\mathbf{J})$ [s] \\
\hline
\software{FEniCS}, automatic code generation, AD & 5.3 \\
\software{ngsolve}, automatic differentiation & 5.9 \\
\software{oomph-lib}, Jacobian symbolically derived by hand & 7.4 \\
\textsl{our code}, Jacobian symbolically derived automatically & 10.8 \\
\textsl{our code}, derivatives by FD & 74.7 \\
\hline
\end{tabular}

\caption{Comparison of assembly times for one residual-Jacobian combination of the \pyth{elastic_single_layer_interface} test case of oomph-lib, but with $300\times 300$ elements. For a static mesh case, we compared a simple Poisson equation, i.e. the example \pyth{two_d_poisson}, but with $1000\times 1000$ elements.}
\label{si:tab:speedcompare}
\end{table}

For multi-physics problems on a moving mesh, our generated code usually outperforms the native \software{C++} implementation of \software{oomph-lib} by a factor of up to $2$ or even more. This performance gain can be attributed to several facts: First of all, for coupled multi-physics equations on a single domain, \software{oomph-lib} calculates the Eulerian derivatives of the shape functions, $\sqrt{\det \mathbf{g}}$ and further quantities multiple times, i.e. once for each of the coupled equations, which does not happen in our dynamically generated code. A lot of performance is also gained by the symbolical derivation with respect to the mesh coordinates for moving meshes. Here, \software{oomph-lib} uses finite differences by default, unless implemented by hand for each equation specifically. Our code generation also automatically plugs in the numerical numbers of fixed parameters before compilation, whereas the compiled \software{oomph-lib} implementation must allow them to vary, i.e. keep all of these as variable parameters. 

When all these factors are eliminated by implementing coupled multi-physics equations in a single equation class in \software{oomph-lib} and using symbolical derivatives with respect to the mesh coordinates, our code suffers from a bit from the overhead for calling the generated code and passing all required information from the core. For a rather simple problem, namely just a Poisson equation, this overhead becomes noticeable, cf. table~\ref{si:tab:speedcompare}. Our code also requires more memory, since e.g. all equations are defined on moving meshes by default, whereas in \software{oomph-lib}, it only allocates e.g. history values for time derivatives of the mesh coordinates if the mesh is indeed moving. This additional memory requirement, however, usually is small compared to the huge memory requirements of linear solvers, in particular direct solvers. \software{oomph-lib} also allows for e.g. a spine-based implementation of a moving mesh, where the entire mesh moves according to degrees of freedom that are only required at e.g. a free surface. This flexibility, which reduces the number of degrees of freedom, is not part of our code yet. For a simple Poisson case, we also compare to \software{FEniCS} \cite{Logg2012}, which also has automatically generated code, subsequently compiled, and utilizes automatic differentiation. Also, we compared against the automatic differentiation of \software{ngsolve} \cite{Schoberl2014}. For the simple Poisson case, the overhead of \software{oomph-lib}'s elemental assembly including plenty of virtual methods and inheritances is definitely visible compared to \software{FEniCS} and \software{ngsolve}. This overhead is even more dominant in our code, where the data has to be passed once more forth and back between the generated code and the \software{oomph-lib} core. \software{nutils} \cite{vanZwieten2022} is based automatic differentiation entirely implemented in \software{python}. The intense performance increase by using high performant \software{C}/\software{C++} codes is remarkable.

\section{Weak formulations and finite element discretizations}
For the weak formulations, we use the following shorthand notations for scalar ($a$, $b$), vectorial ($\vec{a}$, $\vec{b}$) and tensorial ($\mathbf{A}$, $\mathbf{B}$) quantities:
\begin{align}
\weak{a}{b}&=\int_\Omega ab\: {\rm d}\Omega \quad &\iweak{a}{b}&=\int_\Gamma ab \:{\rm d}\Gamma \quad &\iiweak{a}{b}&=\int_{\partial\Gamma} ab \:{\rm d}\partial\Gamma  \nonumber \\
\weak{\vec{a}}{\vec{b}}&=\int_\Omega \vec{a}\cdot\vec{b} \:{\rm d}\Omega \quad &\iweak{\vec{a}}{\vec{b}}&=\int_\Gamma \vec{a}\cdot\vec{b} \:{\rm d}\Gamma \quad &\iiweak{\vec{a}}{\vec{b}}&=\int_{\partial\Gamma} \vec{a}\cdot\vec{b} \:{\rm d}\partial\Gamma \\
\weak{\mathbf{A}}{\mathbf{B}}&=\int_\Omega \vec{A}:\vec{B} \:{\rm d}\Omega   \quad & \iweak{\mathbf{A}}{\mathbf{B}}&=\int_\Gamma \vec{A}:\vec{B} \:{\rm d}\Gamma \quad & \iiweak{\mathbf{A}}{\mathbf{B}}&=\int_{\partial\Gamma} \vec{A}:\vec{B} \:{\rm d}\partial\Gamma \nonumber
\end{align}
By default, the spatial integrations and derivatives are carried out with respect to the underlying coordinates system. For moving meshes, we usually use the Laplace-smoothed mesh implementation, which we solve on a Cartesian coordinate system, since the mesh dynamics is happening e.g. in a two-dimensional projection, not in e.g. the full axisymmetric framework. For a Laplace-smoothed mesh, the integrations and derivatives are furthermore carried out with respect to the a corresponding Lagrangian domain. This is given by the Lagrangian coordinates $\vec{\xi}$, which remain fixed and are initialized with the initial Eulerian position of each node. The residual of a Laplace-smoothed mesh smooths the displacement $\vec{X}-\vec{\xi}$ and its residual contribution hence reads:
\begin{align}
\mathcal{R}_{\vec{X}}^\text{Laplace}=\weak{\nabla_{\vec{\xi}}^\text{cart}(\vec{X}-\vec{\xi})}{\nabla_{\vec{\xi}}^\text{cart}\vec{Y}}_{\xi}^\text{cart}
\end{align}
Here $\vec{X}$ is the field spanned by the Eulerian mesh coordinates and $\vec{Y}$ is the corresponding test function.

\subsection{Fold bifurcation of a detaching hanging droplet}
\subsubsection{Full bulk implementation based on Stokes flow}
For the implementation with the full bulk flow dynamics, we solve for the velocity $\vec{u}$ and the pressure $p$ with corresponding test functions $\vec{v}$ and $q$, respectively. As discretization second/first order Taylor-Hood elements are used. The kinematic boundary condition is enforced by a field of Lagrange multipliers $\lambda_\text{kbc}$ with test function $\mu_\text{kbc}$ of second order at the free surface. The surface tension of unity is imposed via the divergence of the velocity test function. The radial mesh position at the contact line is enforced to be at $\rcontactline$ by a single Lagrange multiplier $\lambda_\text{cl}$ (with test function $\mu_\text{cl}$) that acts on the kinematic boundary condition at the contact line. Thereby, we can have a no-slip condition at the entire solid contact at the top. One could also fix the contact line to $\rcontactline$ strongly, but with a Lagrange multiplier, it is better suited for bifurcation tracking or continuation. Eventually, a final Lagrange multiplier $\lambda_V$ enforces the volume of unity by adjusting the pressure. The volume integral can be cast to an interface integral by virtue of the divergence theorem.

The total weak formulation hence reads:
\begin{align}
\mathcal{R}&=\weak{\nabla\mathbf{u}+\nabla\mathbf{u}^\text{T}-p\mathbf{1}}{\nabla\vec{v}}+\weak{\operatorname{Bo}\vec{e}_z}{\vec{v}}+\weak{\nabla\cdot \vec{u}}{q} +\mathcal{R}_{\vec{X}}^\text{Laplace} \nonumber\\
&+\iweak{(\partial_t \vec{X}-\vec{u})\cdot\unitnormal}{\mu_\text{kbc}}_\text{free surf}+\iweak{\lambda_\text{kbc} }{\vec{Y}\cdot\unitnormal}_\text{free surf}+\iweak{1 }{\nabla\cdot\vec{v}}_\text{free surf} \nonumber\\
&+\iiweak{\vec{X}\cdot\vec{e}_r-\rcontactline}{\mu_\text{cl}}_{cl}+\iiweak{\lambda_\text{cl}}{\mu_\text{kbc}}_{cl} \nonumber\\
&+\iweak{\frac{1}{3}\vec{X}\cdot\unitnormal}{\mu_V}_{\text{free}}-\mu_V+ \weak{\lambda_V}{q} \quad=\quad 0 \label{si:eq:foldbulkweak}
\end{align}
Besides the no-slip boundary condition at the top wall, the radial velocity and the radial mesh coordinates are strongly set to zero at the axis of symmetry. The axial mesh coordinates are pinned to zero at the top wall.

\subsubsection{Interface-only implementation with the Young-Laplace equation}
The alternative implementation employing the Young-Laplace equation only, i.e. not the bulk dynamics, is solved on a line mesh, which is blend in an axisymmetric cylindrical coordinate system to form the shape of the hanging droplet. The domain $\Omega$ is hence a one-dimensional manifold with codimension $1$ and hence all weak terms $\weak{.}{.}=2\pi\int \ldots r {\rm d}l$ are integrals along the moving curved line. The discontinuous elemental normal $\unitnormal$ is smoothed by projection to a second order field $\projectednormal$ (with test function $\vec{m}$). Due to the projection, $\projectednormal$ might not have a unit length, so after normalization, the curvature $\kappa$ (second order as well, test function $\chi$) is calculated by projection of $-\nabla_S\cdot\projectednormal$. The mesh coordinates $\vec{X}$ are moved in normal direction until the fulfill the Young-Laplace equation, again with a Lagrange $\lambda_V$ that acts as additional pressure to fulfill the volume constraint as in the implementation with Stokes flow in the bulk. The mesh coordinates can still move arbitrarily in tangential direction. Here, we use the one-dimensional Lagrangian coordinate $\xi$ as normalized arclength and position all nodes tangentially that way, so that the normalized arclength agrees with the initial normalized arclength before bending the mesh, i.e. with $\xi$.
The normalized arclength coordinate $s$, starting from $s=0$ at the contact line to $s=1$ at the lowest point at the symmetry axis, depends on the entire shape of the droplet. Therefore, a Laplace equation along the mesh with $s=0$ and $s=1$ as Dirichlet conditions at the end points is solved for $s$ with test function $\sigma$, which automatically gives the correct normalized arclength, provided $s$ is solved in a Cartesian coordinate system along the curved line. The mesh coordinates are therefore moved until $s-\xi=0$ holds everywhere. Finally, the radial contact line position is set to $\rcontactline$ by a single Lagrange multiplier $\lambda_\text{cl}$ as before in \eqref{si:eq:foldbulkweak} with a strongly set $z=0$ position, whereas the other end is fixed at $r=0$ by a Dirichlet condition with free $z$-coordinate.

In total, the weak residual form reads
\begin{align}
\mathcal{R}&=\weak{\projectednormal-\unitnormal}{\vec{m}}^\text{cart}+\weak{\kappa+\nabla_S\cdot\left(\frac{\projectednormal}{\|\projectednormal\|}\right)}{\chi}
+\weak{\kappa+\operatorname{Bo}z -\lambda_V}{\unitnormal\cdot\vec{Y}} \nonumber \\
&+\weak{\frac{1}{3}\vec{X}\cdot\unitnormal}{\mu_V}-\mu_V+\weak{\nabla_S^\text{cart} s}{\nabla_S^\text{cart}\sigma}^\text{cart}+\weak{s-\xi}{\vec{t}\cdot\vec{Y}} \nonumber\\
&+\iweak{\vec{X}\cdot\vec{e}_r-\rcontactline}{\mu_\text{cl}}_{cl}+\iiweak{\lambda_\text{cl}}{\mu_\text{kbc}}_{cl}  \quad=\quad 0 \label{si:eq:foldYLweak}
\end{align}

\subsection{Bubble in Hele-Shaw cell with a centered constriction}
For this case, only the pressure $p$ (with test function $q$, second order basis functions) in the outer liquid phase is solved, whereas the bubble is just a hole in the moving mesh. The Neumann condition for the pressure is trivially implemented. The kinematic boundary condition is also just a Neumann contribution at the bubble interface. The dynamic boundary condition is solved by adjusting the mesh positions in normal direction. To that end, a field of Lagrange multipliers $\lambda_\text{dynbc}$ with test function $\mu_\text{dynbc}$ is introduced at the bubble interface. As before in \eqref{si:eq:foldYLweak}, the elemental normal $\unitnormal$ is projected to a continuous normal $\projectednormal$ and the curvature $\kappa$ is calculated from $\projectednormal$.
The unknown bubble pressure $p_\text{B}$ and velocity $U$ are global degrees of freedom -- associated with test functions $q_\text{B}$ and $W$, respectively -- and the corresponding integral constraints are written as integrals over the bubble interface by virtue of the divergence theorem.
In total, this gives the following weak form, together with $p=0$ at $x=L$, where all occurring integrals and derivatives are carried out in a 2d Cartesian system:
\begin{align}
\mathcal{R}&=\weak{b^3\nabla p}{\nabla q}-\iweak{b^3}{q}_{x=-L}+\iweak{b(U e_x+\dot{\vec{X}})}{\unitnormal q }_\text{bubble} \nonumber\\
&+\mathcal{R}_{\vec{X}}^\text{Laplace}+\iweak{\projectednormal-\unitnormal}{\vec{m}}_\text{bubble}+\iweak{\kappa+\nabla_S\cdot\left(\frac{\projectednormal}{\|\projectednormal\|}\right)}{\chi}_\text{bubble} \nonumber \\
&+\iweak{p-p_\text{B}+\frac{1}{3\alpha Q}\left(\frac{1}{b}+\frac{\kappa}{\alpha}\right)}{\mu_\text{dynbc}}_\text{bubble}+\iweak{\lambda_\text{dynbc}}{\unitnormal\cdot\vec{Y}}_\text{bubble} \nonumber\\
&+\iweak{\frac{1}{2}x^2\unitnormal\cdot \vec{e}_x}{W}_\text{bubble} -\iweak{\frac{1}{2}b \vec{x}\cdot\unitnormal}{q_\text{B}}_\text{bubble}-Vq_\text{B} \quad=\quad 0 \, . \label{si:eq:bubbleweak}
\end{align}

\subsection{Onset of convection in a cylindrical Rayleigh-B\'enard system}

For the cylindrical Rayleigh-B\'enard convection problem, the axisymmetric $\resfunc$ and the auxiliary m-dependent ${\resfunc}^m$ residuals are defined. The former includes the nondimensionalized weak forms of: the Navier-Stokes equation with buoyancy bulk term, a zero value volume-averaged constraint for the pressure and the advection-diffusion equation for temperature. Boundary conditions include no-slip in all walls, $T^0=1$ at $z=0$ and $T^0=0$ at $z=1$, adiabatic sidewalls (which are implicitly imposed by integration by parts of the diffusion term in the equation for the temperature). At $r=0$, temperature and pressure must fulfill $\partial_r T^0=0$ and $\partial_r p^0=0$, while for the velocity: $u_r^0=u_\aziangle^0=0$ and $\partial_r u_z^0=0$. The differentiation and integration of the weak forms are carried out in axisymmetric coordinates, but with consideration of a velocity component $u_\aziangle^0$, which would allow e.g. for rotation of the entire system. All fields depend only on $r$, $z$ and $t$, i.e. are independent of the azimuthal angle $\aziangle$ to allow only for axial symmetric solutions.  The axisymmetric residual formulation reads:
\begin{align}
\resfunc&=\weak{\nabla^0\cdot\vec{u}^0}{q}+\weak{\partial_t \vec{u}^0}{\vec{v}} + \weak{\vec{u}^0 \cdot \nabla^0 \vec{u}^0}{\vec{v}} +\weak{Pr(\nabla^0 \vec{u}^0+(\nabla^0 \vec{u}^0)^T)}{\nabla^0 \vec{v}} \nonumber \\
&-\weak{\operatorname{Pr}\operatorname{Ra} p^0}{\nabla^0\cdot\vec{v}}-\weak{\operatorname{Pr}\operatorname{Ra}T^0\vec{e}_z}{\vec{v}} \nonumber\\
&+\weak{\partial_t T^0}{\eta}+\weak{\vec{u}^0 \cdot \nabla^0 T^0}{\eta} +\weak{\nabla^0 T^0}{\nabla^0 \eta} \nonumber \\
&+\weak{p^0}{Q}+\weak{P^0}{q}\, ,
\end{align}
\noindent where $\vec{v}=(v_r,v_\aziangle,v_z)$ is the test function of the velocity, $q$ is the test function of the pressure and $\eta$ is the test function of the temperature. $\nabla^0$ denotes the del operator in cylindrical coordinates, but with $\partial_\aziangle \ldots=0$. The single degree of freedom $P^0$, with corresponding test value $Q$, it the  Lagrange multiplier fixing the average pressure to zero to remove the nullspace of the pressure.

From this entered axisymmetric form, the framework automatically generates the $m$-dependent complex-valued auxiliary residual form by linearization around the axisymmetric solution and considering the $\aziangle$-derivatives applied on the azimuthal modes ($e^{im\aziangle}$ for fields and $e^{-im\aziangle}$ for test functions). Due to the presence of nonlinear terms in Navier-Stokes and advection-diffusion equations, there will be a coupling between axisymmetric stationary solution and the azimuthal perturbations and multiple additional $m$-dependent terms arise due to the $\aziangle$-derivatives, which must be considered now:
\begin{align}
{\resfunc}^m&=\weak{\nabla^0\cdot\vec{u}^m+\frac{im}{r}u_\aziangle^m}{q}+\weak{\partial_t \vec{u}^m}{\vec{v}} \nonumber\\
&+\weak{\vec{u}^0 \cdot \nabla^0 \vec{u}^m+\frac{im}{r}u_\aziangle^0 \vec{u}^m}{\vec{v}} + \weak{\vec{u}^m \cdot \nabla^0 \vec{u}^0+\frac{im}{r}u_\aziangle^m \vec{u}^0}{\vec{v}} \nonumber \\
&+\weak{\operatorname{Pr}(\nabla^0 \vec{u}^m+(\nabla^0 \vec{u}^m)^T)}{\nabla^0 \vec{v}} \nonumber \\
&+\weak{\operatorname{Pr}\left(\frac{m^2u_r^m+3imu_\aziangle^m}{r^2}-\frac{im}{r} \partial_r u_\aziangle^m\right)}{v_r}+\weak{\operatorname{Pr}\frac{2m^2u^m_\aziangle-3imu_r^m}{r^2}}{v_\aziangle}\nonumber\\
&+\weak{\operatorname{Pr}\frac{im}{r} u_r^m}{\partial_r v_\aziangle}+\weak{\operatorname{Pr}\frac{im}{r} u_z^m}{\partial_z v_\aziangle}+\weak{\operatorname{Pr}\left(\frac{m^2u^m_z}{r^2}-\frac{im}{r} \partial_z u_\aziangle^m\right)}{v_z} \nonumber\\
&-\weak{\operatorname{Pr}\operatorname{Ra} p^m}{\nabla^0\cdot\vec{v}-\frac{im}{r}v_\aziangle}-\weak{\operatorname{Pr}\operatorname{Ra}T^m\vec{e}_z}{\vec{v}} \nonumber\\
&+\weak{\partial_t T^m}{\eta}+\weak{\vec{u}^0 \cdot \nabla^0 T^m+\frac{im}{r}u_\aziangle^0 T^m}{\eta}+\weak{\vec{u}^m \cdot \nabla^0 T^0 +\frac{im}{r}u_\aziangle^m T^0 }{\eta} \nonumber\\
&+\weak{\nabla^0 T^m}{\nabla^0 \eta}+\weak{\frac{m^2}{r^2}T^m}{ \eta} +\weak{p^m}{Q}+\weak{P^m}{q}\, ,  \label{eq:weakformRBazi}
\end{align}
For the auxiliary residuals ${\resfunc}^m$, the boundary conditions at $r=0$ differ depending on $m$, as described in the main article. For $m\neq 0$, the perturbation $P^m$ of the  Lagrange multiplier is removed from the system, since the $e^{im\aziangle}$-rotation of the pressure pertubation $p^m$ automatically has a vanishing average in the three-dimensional cylinder.

The Navier-Stokes equation is discretized using Taylor-Hood elements, while the temperature is discretized with first order basis functions.

\subsubsection{A note on the non-integrable singularity in the viscous term}
In \eqref{eq:weakformRBazi}, there are integral contributions proportional to $1/r^2$, which are -- even with the factor $2\pi r$ -- mathematically problematic. For $m=0$, these terms vanish and for $|m|\geq 2$, the strong requirement of $\mathbf{u}^m=0$ and $\mathbf{v}^m=0$ at $r=0$ also removes this singularity. For $|m|=1$, however, $u^m_r$ and $u^m_\aziangle$ are allowed to have nonzero values and, due to the then required Neumann condition $\partial_r u^m_r=\partial_r u^m_\aziangle=0$, also the corresponding test functions $v_r$ and $v_\aziangle$ do not necessarily vanish at $r=0$. These weak contributions hence appear to be singular and non-integrable, but by virtue of the continuity equation, they are in fact not. With the continuity equation, one can rewrite these terms as shown in Refs. \cite{Gelfgat1999,Babor2023} to an integrable formulation. 

\begin{figure}[ht]
\centering
\includegraphics[width=0.5\textwidth]{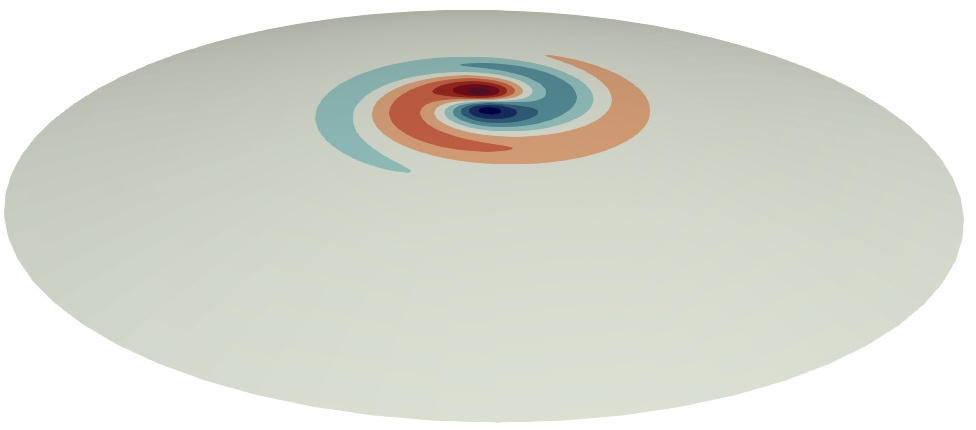}
\caption{Eigenmode of the temperature field of the azimuthal thermal Marangoni instability ($m=1$) of a non-volatile droplet on a heated substrate, which perfectly agrees with Fig.~ 12(a) of Ref.~\cite{Babor2023}. In their work, they obtained a critical Marangoni number of $122.0$, our bifurcation tracking yields $121.96$. This indicates that the explicit treatment of the viscous term for $m=1$ seems to be redundant.}
\label{fig:droplet_babor}
\end{figure}

By numerical experiments, however, it turns out that the discretized $1/r^2$-terms actually do not pose a considerable problem in terms of accuracy. We applied our method to the setting of a non-volatile droplet on a heated or cooled substrate as described by Babor \& Kuhlmann \cite{Babor2023}. In their work, the explicitly reformulate the singularity for $m=1$, whereas in our approach, the discretized $1/r^2$-terms are kept. As shown in Fig.~\ref{fig:droplet_babor}, we get the same results as in Fig.~12 of Ref.~\cite{Babor2023}. Therefore, the explicit treatment of the singularity seems to be redundant. The discretized continuity equation actually automatically enforces a constraint on the azimuthal velocity perturbation so that the apparent singular contribution is numerically well treated. The same observation was also confirmed by private communication with the first author of Ref.~\cite{Babor2023}, who also obtained the same results without the reformulation on the basis of the continuity equation. A rigorous mathematical analysis of the behaviour of the discretized $1/r^2$-terms is beyond the scope of this article.

\section{Example codes}
In this section, we show a few example codes to illustrate the simplicity to define arbitrary problems and equations directly in \software{python}.
Only the equation system has to be assembled, combined with meshes and the system parameters. This simple approach was inspired by the powerful tool \software{FEniCS}, which unfortunately lacks a bit in the required monolithic moving mesh and multi-domain support \software{oomph-lib} has to offer. Therefore, we developed our own framework \software{pyoomph}, that allows easy combination of multiple equations on coupled domains and moving meshes. 

With a few lines of \software{python} code, you can express rather arbitrary multi-physics problems and obtain stationary and transient solutions, and -- as mainly discussed in this work -- perform stability analysis and bifurcation tracking. This section mainly shows how problems and equations can be defined in a concise way directly in \software{python}, it is not considered as full documention of our code. The required \software{python} module called \software{pyoomph} can be made available on request, including a comprehesive documentation with examples. Distribution as free open-source software is envisioned in near future.

\subsection{Example 1: Fold bifurcation of a detaching droplet}
For the detaching droplet, we have developed two different approaches, one with the full bulk equations and the solution of the Young-Laplace equation at the interface only. Since both approaches have the parameters and the output in common, we define a generic problem class, that is the basis for both implementations:
\myinputpython{droplet_detach.py}{1}{20}

\subsubsection{Full bulk implementation based on Stokes flow}
For the full bulk implementation, we import first some implemented equations, then define the problem. The default coordinate system is set to axisymmetric cylindrical coordinates and a mesh is added to the problem (the mesh class itself is skipped here for brevity). Subsequently, the equation system -- consisting of bulk and interface equations as well as boundary conditions and constraints -- is added to the problem:
\myinputpython{droplet_detach.py}{46}{88}

\subsubsection{Interface-only Young-Laplace implementation}
For the problem class based on the Young-Laplace equation, first the corresponding equation class has to be implemented. Within this class, we define all required finite element fields on different spaces. Subsequently, the weak formulation is added to the equation class:
\myinputpython{droplet_detach.py}{89}{131}
The corresponding problem class can now just use this equation and combine it with boundary conditions. We use a one-dimensional line mesh and bend it to a spherical cap shape of a hanging droplet in absence of gravity. Suitable initial and boundary conditions are applied and the assembled equation system is merged with the added mesh:
\myinputpython{droplet_detach.py}{134}{173}

\subsubsection{Obtaining the fold bifurcation curve}
For both problem classes, the fold bifurcation curve is obtained exactly the same way. We first tell that we require the code for a symbolical Hessian. This will write and compile a \software{C} code to fill the Hessian (beside the always generated code for the residual, Jacobian and parameter derivatives thereof). We start by the stationary solution at $\operatorname{Bo}=0$ and use continuation to reach a stationary solution corresponding to a reasonable guess of the critical Bond number. We make a new mesh, solve again on this new mesh and get the eigenvector corresponding to the eigenvalue with real part closest to zero. Then, fold tracking is activated and the augmented fold tracking system is solved. By arclength continuation in the radius, the entire curve can be scanned.
\myinputpython{droplet_detach.py}{176}{199}

\subsection{Example 2: Onset of convection in a cylindrical Rayleigh-B\'enard system}

For the cylindrical Rayleigh-B\'enard convection problem, we define a problem class, which contains the driver codes for the mesh generation and assembly of the equation tree. Additionally, we add a single line of code that augments the system in order to solve for the azimuthal stability. The bifurcation curve is obtained by solving the augmented system as a function of the $Ra$ parameter for an initial aspect ratio $\Gamma$ and subsequently increasing $\Gamma$ by arclength continuation.

\subsubsection{Problem definition}

The problem definition starts by declaring the important parameters, such as the aspect ratio $\Gamma$, the Rayleigh number $Ra$ (and the Prandtl number $Pr$, which is fixed to 1 in this problem). An axisymmetric coordinate system is set so that the differential and integral calculations are performed accordingly and to make it possible to obtain the augmented system. By setting the spatial scale of \pyth{"coordinate_x"} in this coordinate system, we effectively scale the radial coordinate $r \rightarrow \Gamma r$, so that we can modify the cylinder radius without changing the mesh at all, allowing to mesh a simple square geometry.

\myinputpython{RB_problem.py}{1}{23}

With the geometry defined, we can add the equations that define the model. Those are the Navier-Stokes with a body force given by the nondimensional temperature, which in its turn is solved by an advection-diffusion equation. With \pyth{pressure_factor}, we scale the pressure with the $Pr Ra$. This product is entering the bulk force, i.e. the buoyancy. When scaling the pressure the same way, the stationary pressure field is independent of $Pr Ra$. Thereby, one can solve the stationary conductive solution (mainly pressure and temperature fields) for any $Ra$ and change $Ra$ afterwards. Furthermore, we have to fix the null space of the pressure due to the fact that only no-slip boundary conditions are used on a closed domain. Therefore, we enforce the volume-averaged pressure to be zero. No-slip boundary conditions are added on the walls and the temperature is fixed to 1 at the bottom and 0 at the top. By default, the sidewalls are adiabatic. The equations are then added to the discretized domain.

\myinputpython{RB_problem.py}{25}{48}

\subsubsection{Obtaining the azimuthal bifurcation curve}

The driver code for obtaining the azimuthal bifurcation curve is very straightforward when the problem is set. By a single line of code, we let our framework symbolically derive the azimuthal auxiliary residual and from that the corresponding azimuthal mass and Jacobian matrices. For the azimuthal bifurcation tracking, we also derive the required second order derivatives. For each of the $m$ values to be calculated, we first set initial $\Gamma$ and $Ra$ values and solve the system under those conditions. Then, we solve the eigenproblem continuously in order to tweak $Ra$ until $\eigenval_r \approx 0$. This will provide an initial guess for the critical eigenvector in the augmented system. Finally, we increase $\Gamma$ by arclength continuation, while registering the solution of the augmented system at each step and writing the curves $\Gamma - Ra_c$ at each $m$ value into a \pyth{.txt} file.

\myinputpython{RB_tracking.py}{0}{46}

\clearpage
\pagebreak
\bibliographystyle{elsarticle-num} 
\bibliography{submission}

\begin{thebibliography}{10}
\expandafter\ifx\csname url\endcsname\relax
  \def\url#1{\texttt{#1}}\fi
\expandafter\ifx\csname urlprefix\endcsname\relax\def\urlprefix{URL }\fi
\expandafter\ifx\csname href\endcsname\relax
  \def\href#1#2{#2} \def\path#1{#1}\fi

\bibitem{Cross1993}
M.~C. Cross, P.~C. Hohenberg, Pattern formation outside of equilibrium, Rev.
  Mod. Phys. 65 (1993) 851--1112.
\newblock \href {https://doi.org/10.1103/RevModPhys.65.851}
  {\path{doi:10.1103/RevModPhys.65.851}}.

\bibitem{Turing1952}
A.~M. Turing, The chemical basis of morphogenesis, Philos. Trans. R. Soc.
  London Ser. B 237~(641) (1952) 37--72.
\newblock \href {https://doi.org/10.1098/rstb.1952.0012}
  {\path{doi:10.1098/rstb.1952.0012}}.

\bibitem{FitzHugh1961}
R.~FitzHugh, Impulses and physiological states in theoretical models of nerve
  membrane, Biophy. J. 1~(6) (1961) 445--466.
\newblock \href {https://doi.org/https://doi.org/10.1016/S0006-3495(61)86902-6}
  {\path{doi:https://doi.org/10.1016/S0006-3495(61)86902-6}}.

\bibitem{Uecker2014}
H.~Uecker, D.~Wetzel, J.~D.~M. Rademacher, {pde2path} - a matlab package for
  continuation and bifurcation in 2d elliptic systems, Numer. Math-Theory ME.
  7~(1) (2014) 58–106.
\newblock \href {https://doi.org/10.1017/S1004897900000295}
  {\path{doi:10.1017/S1004897900000295}}.

\bibitem{Doedel1998}
E.~J. Doedel, B.~Oldeman, Auto-07p: continuation and bifurcation software,
  Montreal, QC: Concordia University Canada (1998).

\bibitem{Heil2006}
M.~Heil, A.~L. Hazel, {oomph-lib - An Object-oriented multi-physics
  finite-element library}, Lect. Notes Comput. Sci. Eng. 53 (2006) 19--49.
\newblock \href {https://doi.org/10.1007/3-540-34596-5{\_}2}
  {\path{doi:10.1007/3-540-34596-5{\_}2}}.

\bibitem{Salinger2005}
A.~G. Salinger, E.~A. Burroughs, R.~P. Pawlowski, E.~T. Phipps, L.~A. Romero,
  Bifurcation tracking algorithms and software for large scale applications,
  Int. J. Bifurc. Chaos Appl. Sci. Eng. 15~(03) (2005) 1015--1032.
\newblock \href {https://doi.org/10.1142/S0218127405012508}
  {\path{doi:10.1142/S0218127405012508}}.

\bibitem{Salinger2014}
H.~A. Dijkstra, F.~W. Wubs, A.~K. Cliffe, E.~Doedel, I.~F. Dragomirescu,
  B.~Eckhardt, A.~Y. Gelfgat, A.~L. Hazel, V.~Lucarini, A.~G. Salinger, et~al.,
  Numerical bifurcation methods and their application to fluid dynamics:
  Analysis beyond simulation, Commun. Comput. Phys. 15~(1) (2014) 1–45.
\newblock \href {https://doi.org/10.4208/cicp.240912.180613a}
  {\path{doi:10.4208/cicp.240912.180613a}}.

\bibitem{Ham2019}
D.~A. Ham, L.~Mitchell, A.~Paganini, F.~Wechsung, Automated shape
  differentiation in the {Unified Form Language}, Struct. Multidiscipl. Optim.
  60~(5) (2019) 1813--1820.
\newblock \href {https://doi.org/10.1007/s00158-019-02281-z}
  {\path{doi:10.1007/s00158-019-02281-z}}.

\bibitem{Gangl2021}
P.~Gangl, K.~Sturm, M.~Neunteufel, J.~Sch{\"o}berl, Fully and semi-automated
  shape differentiation in {NGSolve}, Struct. Multidiscipl. Optim. 63~(3)
  (2021) 1579--1607.
\newblock \href {https://doi.org/10.1007/s00158-020-02742-w}
  {\path{doi:10.1007/s00158-020-02742-w}}.

\bibitem{Kuznetsov1998}
Y.~A. Kuznetsov, Elements of applied bifurcation theory, Springer, 1998.

\bibitem{Cliffe2000}
K.~Cliffe, S.~Spence, A. and~Tavener, The numerical analysis of bifurcation
  problems with application to fluid mechanics, Acta Numerica 9 (2000) 39 --
  131.
\newblock \href {https://doi.org/10.1017/S0962492900000398}
  {\path{doi:10.1017/S0962492900000398}}.

\bibitem{Bindel2014}
D.~Bindel, M.~Friedman, W.~Govaerts, J.~Hughes, Y.~Kuznetsov, Numerical
  computation of bifurcations in large equilibrium systems in {MATLAB}, J.
  Comput. Appl. Math. 261 (2014) 232--248.
\newblock \href {https://doi.org/10.1016/j.cam.2013.10.034}
  {\path{doi:10.1016/j.cam.2013.10.034}}.

\bibitem{Umbria2016}
J.~S. Umbr{\'\i}a, M.~Net, Numerical continuation methods for large-scale
  dissipative dynamical systems, Eur. Phys. J. Spec. Top. 225~(13-14) (2016)
  2465--2486.

\bibitem{Hazel2019}
A.~L. Hazel, Spatial and temporal adaptivity in numerical studies of
  instabilities, with applications to fluid flows, in: A.~Gelfgat (Ed.),
  Computational Modelling of Bifurcations and Instabilities in Fluid Dynamics,
  Springer International Publishing, Cham, 2019, pp. 75--115.
\newblock \href {https://doi.org/10.1007/978-3-319-91494-7{\_}3}
  {\path{doi:10.1007/978-3-319-91494-7{\_}3}}.

\bibitem{Moore1980}
G.~Moore, A.~Spence, The calculation of turning points of nonlinear equations,
  SIAM J. Numer. Anal. 17~(4) (1980) 567--576.

\bibitem{Tavener1992}
S.~J. Tavener, Symmetric and nonsymmetric equilibria of a rod-and-spring model,
  IMA J. Appl. Math. 49~(1) (1992) 73--102.
\newblock \href {https://doi.org/10.1093/imamat/49.1.73}
  {\path{doi:10.1093/imamat/49.1.73}}.

\bibitem{Griewank1983}
A.~Griewank, G.~Reddien, {The Calculation of Hopf Points by a Direct Method},
  {IMA} J. Numer. Anal. 3~(3) (1983) 295--303.
\newblock \href {https://doi.org/10.1093/imanum/3.3.295}
  {\path{doi:10.1093/imanum/3.3.295}}.

\bibitem{Bauer2002}
C.~Bauer, A.~Frink, R.~Kreckel, Introduction to the {GiNaC} framework for
  symbolic computation within the {C++} programming language, J. Symb. Comput.
  33~(1) (2002) 1--12.
\newblock \href {https://doi.org/10.1006/jsco.2001.0494}
  {\path{doi:10.1006/jsco.2001.0494}}.

\bibitem{Logg2012}
A.~Logg, K.-A. Mardal, G.~Wells, Automated solution of differential equations
  by the finite element method: The {FEniCS} book, Vol.~84, Springer Science \&
  Business Media, 2012.

\bibitem{Eggers1997}
J.~Eggers, Nonlinear dynamics and breakup of free-surface flows, Rev. Mod.
  Phys. 69 (1997) 865--930.
\newblock \href {https://doi.org/10.1103/RevModPhys.69.865}
  {\path{doi:10.1103/RevModPhys.69.865}}.

\bibitem{Kumar2020a}
A.~Kumar, M.~R. Gunjan, K.~Jakhar, A.~Thakur, R.~Raj, Unified framework for
  mapping shape and stability of pendant drops including the effect of contact
  angle hysteresis, Colloids Surf. A: Physicochem. Eng. 597 (2020) 124619.
\newblock \href {https://doi.org/10.1016/j.colsurfa.2020.124619}
  {\path{doi:10.1016/j.colsurfa.2020.124619}}.

\bibitem{Kumar2020b}
A.~Kumar, M.~Gunjan, R.~Raj, On the validity of force balance models for
  predicting gravity-induced detachment of pendant drops and bubbles, Physics
  of Fluids 32 (2020) 101703.
\newblock \href {https://doi.org/10.1063/5.0025488}
  {\path{doi:10.1063/5.0025488}}.

\bibitem{Thompson2014}
A.~B. Thompson, A.~Juel, A.~L. Hazel, Multiple finger propagation modes in
  hele-shaw channels of variable depth, J. Fluid Mech. 746 (2014) 123--164.
\newblock \href {https://doi.org/10.1017/jfm.2014.100}
  {\path{doi:10.1017/jfm.2014.100}}.

\bibitem{FrancoGomez2017}
A.~Franco-Gómez, A.~B. Thompson, A.~L. Hazel, A.~Juel, Bubble propagation on a
  rail: a concept for sorting bubbles by size, Soft Matter 13 (2017)
  8684--8697.
\newblock \href {https://doi.org/10.1039/C7SM01478C}
  {\path{doi:10.1039/C7SM01478C}}.

\bibitem{Keeler2019}
J.~S. Keeler, A.~B. Thompson, G.~Lemoult, A.~Juel, A.~L. Hazel, The influence
  of invariant solutions on the transient behaviour of an air bubble in a
  {H}ele-{S}haw channel, Proc. Math. Phys. Eng. Sci. A 475~(2232) (2019)
  20190434.
\newblock \href {https://doi.org/10.1098/rspa.2019.0434}
  {\path{doi:10.1098/rspa.2019.0434}}.

\bibitem{Thompson2021}
A.~B. Thompson, Bifurcations of drops and bubbles propagating in variable-depth
  {H}ele-{S}haw channels, J. Eng. Math. 129~(1) (Jul. 2021).
\newblock \href {https://doi.org/10.1007/s10665-021-10146-y}
  {\path{doi:10.1007/s10665-021-10146-y}}.

\bibitem{Gaillard2021}
A.~Gaillard, J.~S. Keeler, G.~L. Lay, G.~Lemoult, A.~B. Thompson, A.~L. Hazel,
  A.~Juel, The life and fate of a bubble in a geometrically perturbed
  {H}ele-{S}haw channel, J. Fluid Mech. 914 (Mar. 2021).
\newblock \href {https://doi.org/10.1017/jfm.2020.844}
  {\path{doi:10.1017/jfm.2020.844}}.

\bibitem{Yim2016}
E.~Yim, P.~Billant, Analogies and differences between the stability of an
  isolated pancake vortex and a columnar vortex in stratified fluid, J. Fluid
  Mech. 796 (2016) 732–766.
\newblock \href {https://doi.org/10.1017/jfm.2016.248}
  {\path{doi:10.1017/jfm.2016.248}}.

\bibitem{Yim2022}
E.~Yim, A.~Bouillant, D.~Quéré, F.~Gallaire, Leidenfrost flows: instabilities
  and symmetry breakings, Flow 2 (2022) E18.
\newblock \href {https://doi.org/10.1017/flo.2022.5}
  {\path{doi:10.1017/flo.2022.5}}.

\bibitem{Babor2023}
L.~Babor, H.~C. Kuhlmann, Linear stability of thermocapillary flow in a droplet
  attached to a hot or cold substrate, Phys. Rev. Fluids 8 (2023) 114003.
\newblock \href {https://doi.org/10.1103/PhysRevFluids.8.114003}
  {\path{doi:10.1103/PhysRevFluids.8.114003}}.

\bibitem{Liu2021}
H.~Liu, J.~He, Z.~Zeng, Z.~Qiu, Instabilities of thermocapillary-buoyancy flow
  in a rotating annular pool for medium-prandtl-number fluid, Physical Review E
  104 (09 2021).
\newblock \href {https://doi.org/10.1103/PhysRevE.104.035101}
  {\path{doi:10.1103/PhysRevE.104.035101}}.

\bibitem{Batchelor1962}
G.~K. Batchelor, A.~E. Gill, Analysis of the stability of axisymmetric jets, J.
  Fluid Mech. 14~(4) (1962) 529–551.
\newblock \href {https://doi.org/10.1017/S0022112062001421}
  {\path{doi:10.1017/S0022112062001421}}.

\bibitem{Boronska2006}
K.~Borońska, L.~S. Tuckerman, Standing and travelling waves in cylindrical
  {R}ayleigh–{B}énard convection, J. Fluid Mech. 559 (2006) 279–298.
\newblock \href {https://doi.org/10.1017/S0022112006000309}
  {\path{doi:10.1017/S0022112006000309}}.

\bibitem{Gelfgat1999}
A.~Gelfgat, Z.~Bar-Yoseph, A.~Solan, T.~Kowalewski, An axisymmetry-breaking
  instability of axially symmetric natural convection, Int. J. Transport
  Phenomena 1 (1999) 173--190.

\bibitem{Lowry1995}
B.~J. Lowry, P.~H. Steen, Capillary surfaces: Stability from families of
  equilibria with application to the liquid bridge, Proc. Math. Phys. Eng. Sci.
  449~(1937) (1995) 411--439.

\bibitem{Hazel2005}
A.~L. Hazel, M.~Heil, Surface-tension-induced buckling of liquid-lined elastic
  tubes: a model for pulmonary airway closure, Proc. R. Soc. A: Math. Phys.
  Eng. Sci. 461~(2058) (2005) 1847--1868.
\newblock \href {https://doi.org/10.1098/rspa.2005.1453}
  {\path{doi:10.1098/rspa.2005.1453}}.

\bibitem{Bostwick2015}
J.~Bostwick, P.~Steen, Stability of constrained capillary surfaces, Annu. Rev.
  Fluid Mech. 47~(1) (2015) 539--568.
\newblock \href {https://doi.org/10.1146/annurev-fluid-010814-013626}
  {\path{doi:10.1146/annurev-fluid-010814-013626}}.

\bibitem{Slobozhanin1997}
L.~A. Slobozhanin, J.~I.~D. Alexander, A.~H. Resnick, Bifurcation of the
  equilibrium states of a weightless liquid bridge, Physics of Fluids 9~(7)
  (1997) 1893--1905.

\bibitem{Li2019}
Y.~Li, C.~Diddens, A.~Prosperetti, K.~L. Chong, X.~Zhang, D.~Lohse, Bouncing
  oil droplet in a stratified liquid and its sudden death, Phys. Rev. Lett.
  122~(15) (2019) 154502.

\bibitem{Li2021}
Y.~Li, C.~Diddens, A.~Prosperetti, D.~Lohse, Marangoni instability of a drop in
  a stably stratified liquid, Phys. Rev. Lett. 126~(12) (2021) 124502.

\bibitem{Li2022}
Y.~Li, J.~G. Meijer, D.~Lohse, Marangoni instabilities of drops of different
  viscosities in stratified liquids, J. Fluid Mech. 932 (2022) A11.

\bibitem{Meijer2023}
J.~G. Meijer, Y.~Li, C.~Diddens, D.~Lohse, On the rising and sinking motion of
  bouncing oil drops in strongly stratified liquids, J. Fluid Mech. 966 (2023)
  A14.

\bibitem{Herrada2023}
M.~A. Herrada, J.~M. Montanero, L.~Carri\'on, Dynamics of a silicone oil drop
  submerged in a stratified ethanol-water bath, Phys. Rev. E 108 (2023) 065104.
\newblock \href {https://doi.org/10.1103/PhysRevE.108.065104}
  {\path{doi:10.1103/PhysRevE.108.065104}}.

\bibitem{Diddens2017}
C.~Diddens, H.~Tan, P.~Lv, M.~Versluis, J.~Kuerten, X.~Zhang, D.~Lohse,
  Evaporating pure, binary and ternary droplets: thermal effects and axial
  symmetry breaking, J. Fluid Mech. 823 (2017) 470–497.
\newblock \href {https://doi.org/10.1017/jfm.2017.312}
  {\path{doi:10.1017/jfm.2017.312}}.

\bibitem{Sefiane2008}
K.~Sefiane, J.~R. Moffat, O.~K. Matar, R.~V. Craster, {Self-excited
  hydrothermal waves in evaporating sessile drops}, Appl. Phys. Lett. 93~(7)
  (2008) 074103.
\newblock \href {https://doi.org/10.1063/1.2969072}
  {\path{doi:10.1063/1.2969072}}.

\bibitem{Gauthier2019}
A.~Gauthier, C.~Diddens, R.~Proville, D.~Lohse, D.~van Der~Meer,
  Self-propulsion of inverse leidenfrost drops on a cryogenic bath, Proc. Natl.
  Acad. Sci. 116~(4) (2019) 1174--1179.

\bibitem{Maass2016}
C.~C. Maass, C.~Kr\"{u}ger, S.~Herminghaus, C.~Bahr, Swimming droplets, Annu.
  Rev. Condens. Matter Phys. 7~(1) (2016) 171--193.
\newblock \href {https://doi.org/10.1146/annurev-conmatphys-031115-011517}
  {\path{doi:10.1146/annurev-conmatphys-031115-011517}}.

\bibitem{Michelin2023}
S.~Michelin, Self-propulsion of chemically active droplets, Annu. Rev. Fluid
  Mech. 55~(1) (2023) 77--101.
\newblock \href {https://doi.org/10.1146/annurev-fluid-120720-012204}
  {\path{doi:10.1146/annurev-fluid-120720-012204}}.

\bibitem{Edwards2021}
A.~M.~J. Edwards, E.~Ruiz-Guti\'errez, M.~I. Newton, G.~McHale, G.~G. Wells,
  R.~Ledesma-Aguilar, C.~V. Brown, Controlling the breakup of toroidal liquid
  films on solid surfaces, Sci. Rep. 11~(1) (2021) 8120.
\newblock \href {https://doi.org/10.1038/s41598-021-87549-5}
  {\path{doi:10.1038/s41598-021-87549-5}}.

\bibitem{Geuzaine2009}
C.~Geuzaine, J.-F. Remacle, Gmsh: A 3-d finite element mesh generator with
  built-in pre- and post-processing facilities, Int. J. Numer. Methods Eng.
  79~(11) (2009) 1309--1331.
\newblock \href {https://doi.org/https://doi.org/10.1002/nme.2579}
  {\path{doi:https://doi.org/10.1002/nme.2579}}.

\bibitem{Schoberl2014}
J.~Sch{\"o}berl, C++ 11 implementation of finite elements in ngsolve, in: ASC
  Report 30/2014, Vienna University of Technology, 2014, pp. 1--23.

\bibitem{vanZwieten2022}
J.~van Zwieten, G.~van Zwieten, W.~Hoitinga, Nutils 7.0 (2022).
\newblock \href {https://doi.org/10.5281/zenodo.6006701}
  {\path{doi:10.5281/zenodo.6006701}}.

\end{thebibliography}
\end{document}